\documentclass[twocolumn, prb, showpacs]{revtex4-1}
\usepackage{graphicx}
\usepackage{dcolumn}
\usepackage{bm}
\usepackage{color}

\begin{document}
\title{Helical magnetic order and Fermi surface nesting in non-centrosymmetric ScFeGe}

\author{Sunil K. Karna,$^{1,2,\ast}$ D. Tristant,$^{3}$ J. K. Hebert,$^{1,\dagger}$ G. Cao,$^{1,4}$ R. Chapai,$^{1}$ W. A. Phelan,$^{1,\dagger}$ Q. Zhang,$^{1,5}$ Y. Wu,$^{1,5}$ C. Dhital,$^{1,6}$ Y. Li,$^{1}$  H. B. Cao,$^{5}$ W. Tian,$^{5}$ C. R. Dela Cruz,$^{5}$ A. A. Aczel,$^{5,7}$ O. Zaharko,$^{8}$ A. Khasanov,$^{9}$ M. A. McGuire,$^{10}$ A. Roy,$^{11}$ W. Xie,$^{12}$ D. A. Browne$^{1}$ I. Vekhter,$^{1}$ V. Meunier,$^{13}$ W. A. Shelton,$^{3}$  P. W. Adams,$^{1}$  P. T. Sprunger,$^{1,11}$ D. P. Young,$^{1}$ R. Jin,$^{1}$ J. F. DiTusa,$^{1,\ddagger,\ast}$}

\affiliation{$^1$Department of Physics \& Astronomy, Louisiana State University, Baton Rouge, LA 70803, USA}
\affiliation{$^2$Department of Physics \& Center for Materials Research, Norfolk State University, Norfolk, VA 23504, USA}
\affiliation{$^3$Cain Department of Chemical Engineering, Louisiana State University, Baton Rouge, LA 70803, USA}
\affiliation{$^4$Materials Genome Institute and Department of Physics, Shanghai University, Shanghai 200444, China}
\affiliation{$^5$Neutron Scattering Division, Oak Ridge National Laboratory, Oak Ridge, TN 37831, USA}
\affiliation{$^6$Department of Physics, Kennesaw State University, Marietta, GA 30060, USA}
\affiliation{$^7$Department of Physics and Astronomy, University of Tennessee, Knoxville, TN 37996, USA}
\affiliation{$^8$Laboratory for Neutron Scattering and Imaging, Paul Scherrer Institut, Villigen PSI, Switzerland}
\affiliation{$^9$M$\ddot{o}$ssbauer Effect Data Center, University of North Carolina, Asheville, NC 28804, USA}
\affiliation{$^{10}$Materials Science and Technology Division, Oak Ridge National Laboratory, Oak Ridge,TN, 37831, USA}
\affiliation{$^{11}$Center for Advanced Microstructures and Devices (CAMD), Louisiana State University, Baton Rouge, LA}
\affiliation{$^{12}$Department of Chemistry, Louisiana State University, Baton Rouge, LA 70803, USA}
\affiliation{$^{13}$Department of Physics, Applied Physics, and Astronomy, Rensselaer Polytechnic Institute, Troy, NY 12180, USA}

\date{\today}

\begin{abstract}

An investigation of the structural, magnetic, thermodynamic, and charge transport properties of non-centrosymmetric hexagonal ScFeGe reveals it to be an anisotropic metal with a transition to a weak itinerant incommensurate helimagnetic state below $T_N = 36$ K.  Neutron diffraction measurements discovered a temperature and field independent helical wavevector \textbf{\textit{k}} = (0 0 0.193) with magnetic moments of 0.53 $\mu_{B}$ per formula unit confined to the {\it ab}-plane. Density functional theory calculations are consistent with these measurements and find several bands that
cross the Fermi level along the {\it c}-axis with a nearly degenerate set of flat bands just above the Fermi energy. The anisotropy found in the electrical transport is reflected in the calculated Fermi surface, which consists of several warped flat sheets along the $c$-axis with two regions of significant nesting, one of which has a wavevector that closely matches that found in the neutron diffraction. The electronic structure calculations, along with a strong anomaly in the {\it c}-axis conductivity at $T_N$, signal a Fermi surface driven magnetic transition, similar to that found in spin density wave materials. Magnetic fields applied in the {\it ab}-plane result in a metamagnetic transition with a threshold field of $\approx$ 6.7 T along with a sharp, strongly temperature dependent, discontinuity and a change in sign of the magnetoresistance for in-plane currents. Thus, ScFeGe is an ideal system to investigate the effect of in-plane magnetic fields on an easy-plane magnetic system, where the relative strength of the magnetic interactions and anisotropies determine the topology and magnetic structure.
     
\end{abstract}

\maketitle
\section{Introduction}
    The magnetism found in non-centrosymmetric (NCS) magnetic materials is compelling because of the topologically non-trivial nanoscopic magnetic structures discovered that offer potential for new magnetic information manipulation and storage technologies\cite{Robler2006,Muhlbauer2009,Yu2011,Nagaosa2013,Tannous2017,Myers1999,Togawa2012,Karna2019}. For example, MnSi, which has a NCS crystal structure that also lacks mirror symmetry so that it is chiral, was discovered to host a skyrmion lattice, an ordered array of topologically stable knots of spin structure\cite{Ishikawa1976,Ishikawa1977,Pfleiderer2000, Muhlbauer2009}. The cause of this interesting behavior is the Dzyaloshinskii-Moriya (DM) interaction, which occurs only in NCS materials resulting in topologically non-trivial magnetic structures, such as the magnetic \textit{B20} materials, including FeGe, Fe$_{1-x}$Co$_x$Si, and MnGe\cite{Yu2011,Grigoriev2009,Tanigaki2015}. While the cubic symmetry appears to promote skyrmion lattice formation, hexagonal chiral magnetic materials, such as Cr$_{1/3}$NbS$_2$ and Mn$_{1/3}$NbS$_2$, display helimagnetism with a transition to unusual magnetic phases, including the chiral magnetic soliton lattice\cite{Togawa2012,Karna2019}.

Although the DM interaction is expected to be significant in all NCS materials, thus far topologically interesting magnetic phases have not been found in bulk achiral materials outside of those caused by demagnetization fields. The DM interaction stems from spin-orbit coupling and is antisymmetric, favoring perpendicular arrangements of magnetic moments\cite{Koretsune2015}. The role that the DM interaction plays in polar NCS compounds, ones that retain mirror symmetry, is not as well explored. The lack of a left/right asymmetry in the crystal structure of polar NCS systems would tend to rule out topological effects, despite the
presence of the DM interaction. Topological features may, however, be present in polar NCS helimagnets, since the magnetic ordering itself breaks mirror symmetry. The discovery of such features in NCS magnets would also be important because of the interest in magnetic topological structures, such as skyrmions, for possible technological applications (skyrmtronics)\cite{Fert2013,Finocchio2016}.    

Helimagnetic ordering in easy-plane systems has been of high recent interest. The application of a magnetic field along the easy-plane causes a distortion of the helical order leading to interesting magnetic structures, as moments tend to align with the external field. The sequence of magnetic structures that results as the system evolves to the field polarized state depends heavily on the symmetry of the crystal lattice, in particular whether the space group is chiral or achiral, and the relative strength of important interactions. These include the exchange interaction, the DM interaction, the easy-plane anisotropy, and the dipole-dipole interactions through the local demagnetization fields. The most interesting systems are those where the easy-plane anisotropy is sufficient to confine the moments to the {\it ab}-plane, even when exposed to in-plane fields. For systems that are dominated by the DM interaction, such that one chirality or handedness of the helical state is highly preferred, the magnetic structure evolves to one where regions of magnetic moments aligned with {\it H} are separated by $2\pi$ chiral domain walls. This state is known as the magnetic soliton lattice\cite{Togawa2012,Karna2019}. Further increases in {\it H} eventually cause a breakdown of the topological protection afforded this state, so that it transforms to the topologically trivial field-polarized state. 

In cases where the DM interaction is not sufficient to cause a significant energy difference between the two chiralities, such as in achiral NCS magnets, a similar distortion of the helimagnetic structure results with  the application of small in-plane fields. However, a first order metamagnetic transition at $H_{MM} \sim 1/2 H_{sat}$, where $H_{sat}$ is the saturation field,  leads to a magnetic fan state. This transition is usually hysteretic, as a large fraction of magnetic moments rearrange at $H_{MM}$. The fan state consists of alternating regions along the $c$-axis of left- and right-handed twists forming a lattice of kinks, where the helicity reverses and is thus a topologically trivial state. Further increases in field result in a smooth reduction in turn angles between kinks until the system reaches full polarization. Recent investigations have discovered more complex magnetic structures including helifan states and spin-slip phases\cite{Chattopadhyay1994}.  

\begin{figure}
\begin{center}
\includegraphics[width=3.0in]{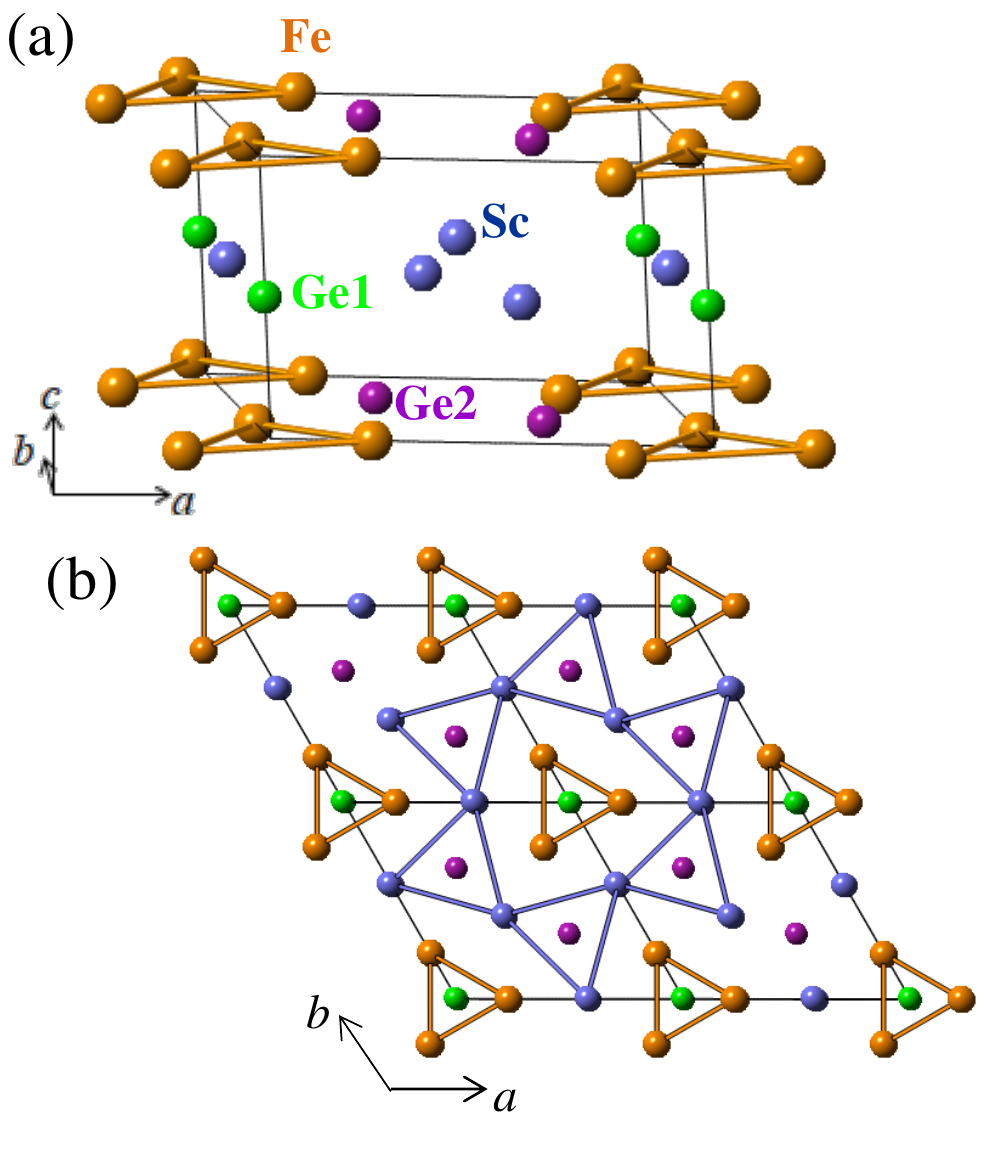}
\end{center}
\caption{\label{fig-crystru} Crystal structure. (a) Crystal structure of ScFeGe, highlighting the triangular network of Fe located at the \textit{3f}-site. (b) Two-dimensional view of the crystal structure along the \textit{c}-axis displaying the distorted Kagome-lattice of Sc.}
\end{figure}

  Here, we explore the physical properties of ScFeGe, which crystallizes in the Fe$_2$P-type [ZrNiAl structure type, symmetry group $P\bar{6}2m$] hexagonal crystal structure that is common among transition metal ternary compounds\cite{Kotur1984}. This crystal structure, shown in Fig.~\ref{fig-crystru}, is NCS, but achiral, and is adopted by a number of magnetic materials, including several that display interesting magnetocaloric properties, such as Fe$_2$P, MnFeP$_{1-x}$Ge$_x$~\cite{Liu2009}, and MnFeP$_{1-x}$As$_x$\cite{Tegus2002}. Those most heavily investigated contain two magnetic species, one on the 3g site that is described as forming a distorted Kagome lattice (the Sc site in Fig.~\ref{fig-crystru}), and one on the 3f site (Fe) that forms well-separated triangular stacks. These materials tend to order ferromagnetically, where the magnetic phase transition is accompanied by a discontinuous and hysteretic change in the c/a lattice constant ratio, making them good candidates for magnetocaloric applications. In contrast, ScFeGe has only a single species that is typically magnetic, Fe, and is therefore a somewhat simpler system to explore the role of crystal symmetry on its magnetic structure and the possibility of topologically interesting spin textures. Very little is known about this compound outside of its crystal structure~\cite{Skolozdra1991}. We discover that ScFeGe is a weak itinerant magnet, ordering into a helimagnetic state with an incommensurate wavevector of \textbf{\textit{k}} = (0 0 0.193) (in units of rlu) at 36 K,
caused by a Fermi surface instability. Despite its itinerant character, we find that ScFeGe has many features in common with rare earth elemental metals, such as Dy, Ho, and Tb.

\begin{figure}
\begin{center}
\includegraphics[width=3.0in]{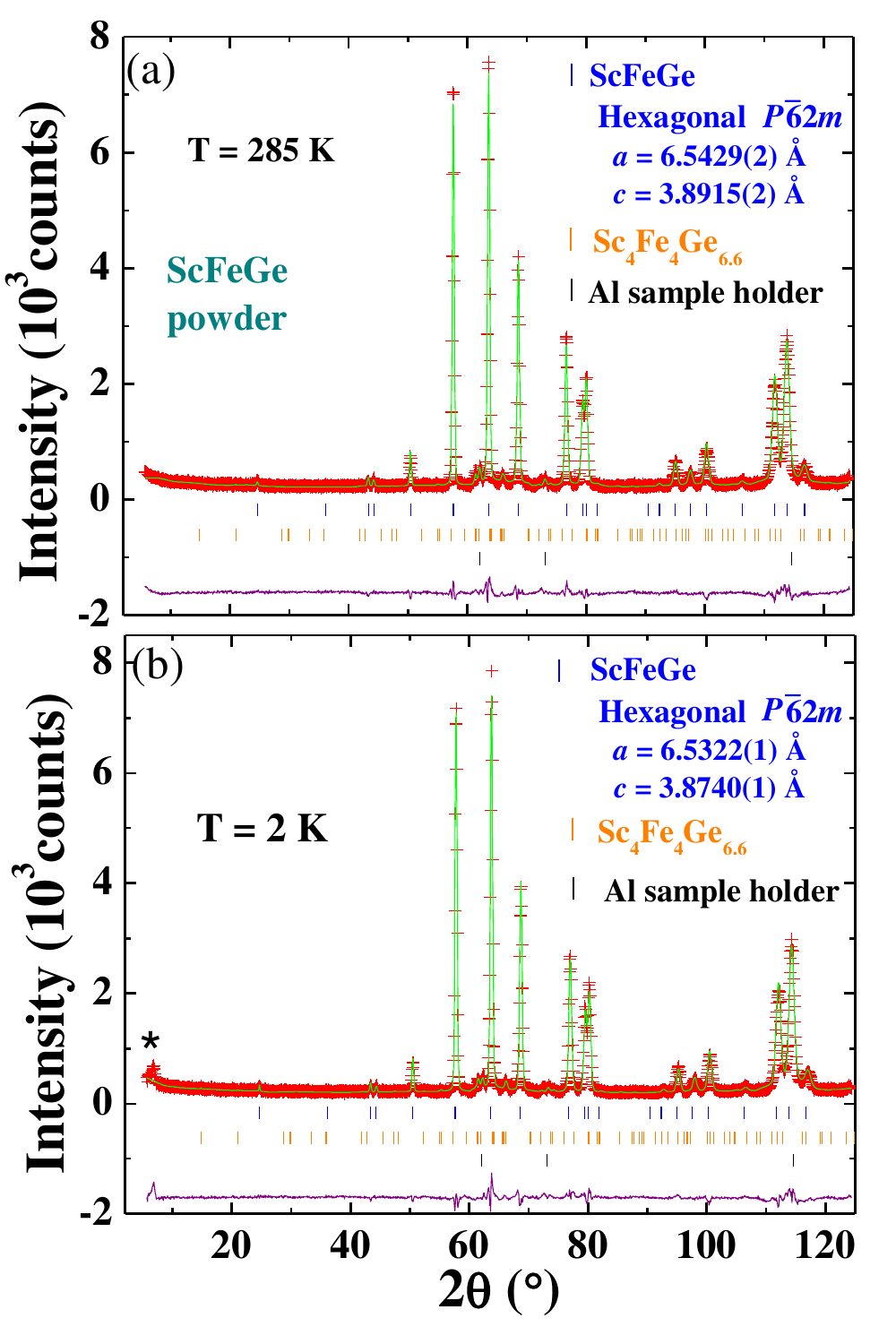}
\end{center}
\caption{\label{NPD} Neutron Powder Diffraction. High-resolution neutron powder-diffraction data (red crosses) taken at (a) 285 K, and (b) 2 K. The green line is the result of a structural refinement which indicated an Fe$_2$P hexagonal crystal structure having the $P\bar{6}2m$ space group with peak positions indicated by the blue vertical lines. Several low intensity peaks were identified as a originating from a small Sc$_4$Fe$_4$Ge$_{6.6}$ impurity phase and from the Al sample holder. The refinement includes these contributions with peak positions indicated by the orange and black vertical lines. The remaining difference between the data and the fit is included as a purple line at the bottom of the figure. The asterisk marked in (b) is a magnetic peak. Data were collected at the HB-2A beam line at ORNL.} 
\end{figure}

\section{Experimental details} 
            
\textbf{Sample Preparation}. 
Polycrystalline samples of ScFeGe were prepared via arc-melting techniques. High purity (99.999 \%) starting materials consisting of the elements Sc, Fe, and Ge were melted on a water-cooled copper hearth in a Zr-gettered ultra-pure argon
atmosphere.  Each button was flipped and melted several times to ensure homogeneity. Only those polycrystalline samples of ScFeGe
corresponding to a 1\% or less calculated mass loss were employed for the single crystal growth. Five single-phase buttons that were produced in this manner were placed in a second copper hearth in the same Zr-gettered ultra-pure argon atmosphere and arc-melted for a second time to produce a rod for crystal growth employing the floating-zone method. Feed and seed rods were obtained from this arc-melted rod. A large single crystal of ScFeGe was synthesized, for the first time, in a floating-zone furnace-a Cannon FZ- MDH20020 2-mirror image furnace with 2 $\times$ 1000 W lamps under Ar gas flow. A growth
rate of 2 mm/h was used to promote low mosaicity, and the seed and feed rods were counter-rotated at 12 to 14 rpm. The obtained cylindrical crystal with a diameter of about 2.5 mm was cut to a suitable size for magnetic, transport, heat capacity, and neutron diffraction measurements.

Powder X-ray diffraction (PXRD) measurements of a crushed single crystal and our polycrystalline samples  were carried out on a PANalytical Empyrean multi-stage X-ray diffractometer with Cu K$\alpha_1$ radiation ($\lambda$= 1.54059 \AA ). The resulting X-ray diffraction patterns were analyzed using the General Structure Analysis System (GSAS) program~\cite{Larson1990} following the Rietveld profile refining method\cite{Rietveld1969}. The refinement confirmed a hexagonal crystal structure (Fig.~\ref{fig-crystru}) with space group $P\bar{6}2m$. The lattice parameters for the crushed single crystal were $a$ = 6.5410(4) \AA \hspace{0.1cm} and $c$ = 3.8905(3) \AA \hspace{0.1cm}(Table S1 in the supplemental material~\cite{Supplement}). For the polycrystalline samples, the lattice constants varied somewhat, being as large as $a$ = 6.5451(4) \AA \hspace{0.1cm} and $c$ = 3.8938(2)\AA \hspace{0.1cm}, with the variation corresponding to increases in the Ge content of the samples (see below). These
measurements were followed by single crystal X-ray diffraction characterization performed on single crystals mounted onto a glass fiber tip using epoxy, attached to a goniometer head via the ends of brass pins, and placed on a Nonius Kappa CCD X-ray diffractometer equipped with Mo-K$\alpha$ radiation ($\lambda$ = 0.71073 \AA). The single crystal diffraction confirmed the crystal structure determined from PXRD and provided orientation of the crystals for further characterization.

Chemical analysis was performed using an electron probe X-ray microanalyzer (EPMA) with a JEOL JSX-8230 analyzer located at the Shared Instrumentation Facility (SIF) at Louisiana State University (LSU). This instrument allows simultaneous measurement via wavelength dispersive spectroscopy (WDS) and energy dispersive spectroscopy (EDS) techniques. EDS and WDS provided a chemical composition of the single crystals as Sc$_{1.00}$Fe$_{1.02(2)}$Ge$_{0.99(2)}$, hereafter referred to as ScFeGe. In addition, the polycrystalline samples were found to typically contain as much as $8\%$ excess Ge, creating a negative chemical pressure as  indicated by slightly increased lattice constants. The excess Ge content and unit cell volume correlated well with an increased magnetic ordering temperature (see below). 
      
\textbf{Magnetic, Charge Transport, and Thermodynamic Measurements}. Magnetization and magnetic susceptibility measurements were carried out on both polycrystalline and single crystal samples in a Quantum Design(QD) Magnetic Property Measurement System (MPMS) superconducting quantum interference device (SQUID) magnetometer with a 7-T superconducting magnet. For the single crystals, a magnetic field was applied either parallel or perpendicular to the crystallographic \textit{c}-axis. The electrical resistivity and magnetoresistance were measured on polished single crystals with contacts formed via conductive epoxy (Epotek H20E) and thin platinum wire. These were standard four-terminal $dc$ resistance measurements with a current of 4 mA applied parallel to the $a$- and $c$-axes of the crystals. Data were collected in a QD Physical Property Measurement System (PPMS) with a 14 T
superconducting magnet. The specific heat capacity was measured using a time-relaxation method in a QD PPMS between 2 and 100 K.

\begin{figure*}
\begin{center}
\includegraphics[width=5.5in]{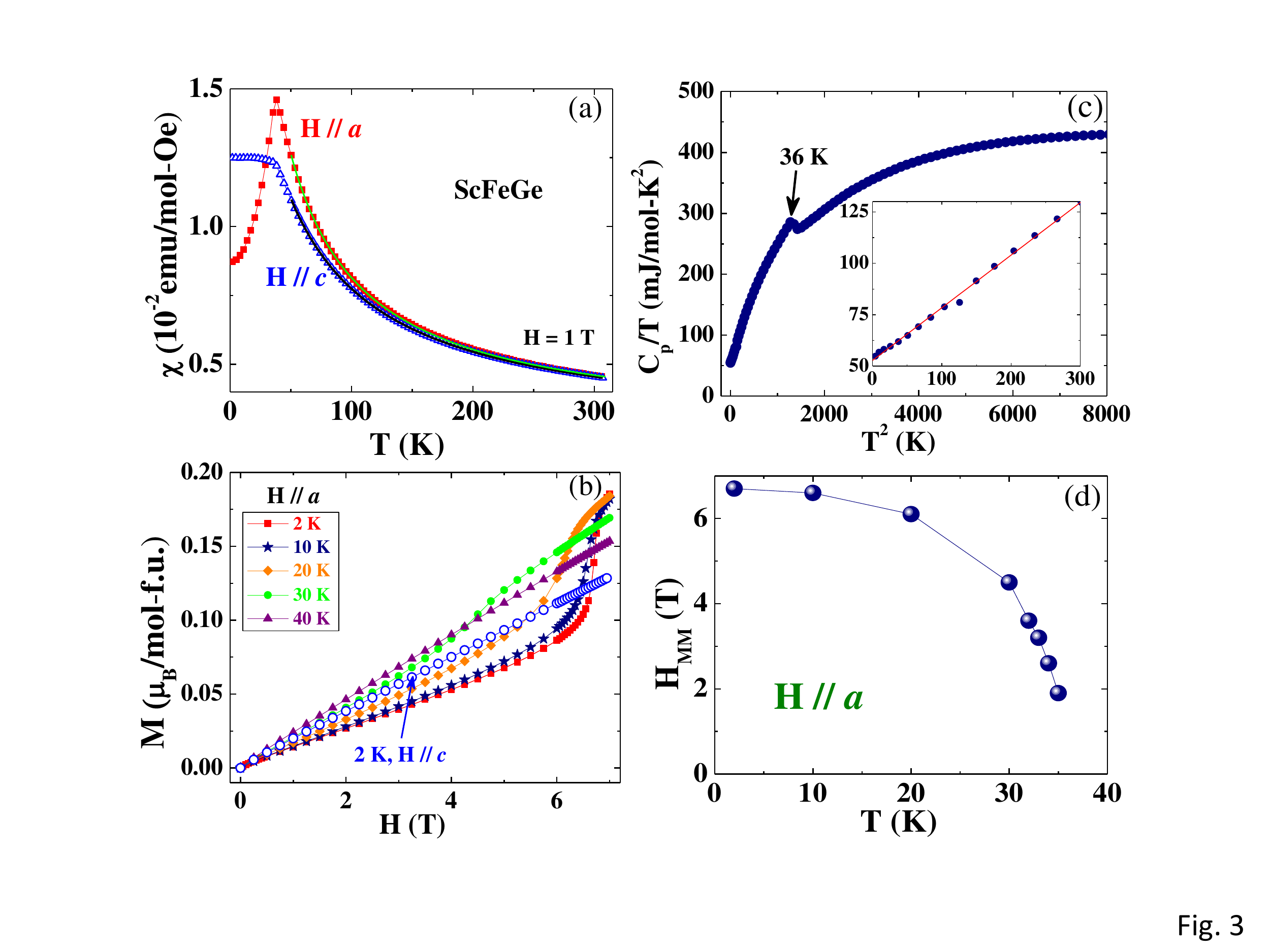}
\end{center}
\caption{\label{fig-chiT} Magnetic susceptibility, magnetization, and specific heat capacity. (a) Temperature, \textit{T}, dependence of the magnetic susceptibility, $\chi$, with magnetic field, $H$, oriented parallel to the \textit{a} and \textit{c} crystallographic axes. (b) \textit{H}-dependence of the magnetization, \textit{M}, for the same field orientations. (c) Specific heat capacity divided by $T$, $C_p(T)/T$, vs. $T^2$ at zero field. Inset: Low temperature region shown on an expanded scale. The solid line is the best fit of a linear form to the data. (d) $T$ dependence of the metamagnetic field, $H_{MM}$, for $H \parallel a$. }
\end{figure*}

\textbf{Neutron Diffraction}. Neutron powder diffraction (NPD) measurements were performed using the HB-2A Neutron Powder Diffractometer at the High Flux Isotope Reactor (HFIR) at Oak Ridge National Laboratory (ORNL). A 5-gram polycrystalline sample of ScFeGe was sealed in an aluminum sample can in a helium atmosphere. The temperature dependence of the diffraction profiles in zero and in an applied magnetic field of 4 T was measured in the temperature range between 2-300 K using a
vertical-field cryomagnet. An incident neutron beam with a wavelength of 2.41 \AA, defined by a Ge (113) crystal monochromator, was employed for these measurements. In addition, the crystal and magnetic structure of a single crystal taken from the floating zone sample were investigated by neutron diffraction measurements using the four-circle diffractometer HB-3A. Single crystal diffraction measurements under applied magnetic field were also performed using a vertical HB-1A triple-axis spectrometer at ORNL using vertical cryomagnet. Data collected at HB-3A included 105 reflections taken at $T = 4$ and 285 K using a wavelength of 1.546 \AA (Si-220 monochromator)\cite{Chakoumakos2011}. Both the powder
and single crystal neutron experiments reproduced the crystal structure determined from our X-ray diffraction. The neutron powder diffraction patterns were analyzed using the GSAS program, whereas single-crystal nuclear and magnetic structural refinements were performed with the FULLPROF suite~\cite{Carvajal1993}.  

Fig.~\ref{NPD}(a) displays the neutron powder diffraction patterns (crosses) that were taken at 285 K along with the results of the refinement (solid lines), with the difference plotted at the bottom. The refinement was conducted assuming a hexagonal symmetry with a space group of $P\bar{6}$2m and taking the pseudo-Voigt function for the peak profiles. The refinement indicated that several low intensity peaks originated from a small impurity phase identified as Sc$_4$Fe$_4$Ge$_{6.6}$ as well as the Al sample holder. The low intensity impurity phase of Sc$_4$Fe$_4$Ge$_{6.6}$ and Al sample holder have also been taken into consideration in the refinement. The solid vertical lines mark the calculated positions of the Bragg reflections of the proposed crystalline structure, the Sc$_4$Fe$_4$Ge$_{6.6}$ impurity phase and the Al sample holder. Table S2 in the Supplemental Material\cite{Supplement} presents the refinement results of ScFeGe obtained at 285 K. No structural change with cooling is detected, as demonstrated by the diffraction pattern obtained at 2 K (Fig.~\ref{NPD}(b)). Neutron diffraction experiments were also performed at the single-crystal neutron diffractometer ZEBRA at the Swiss Neutron Spallation Source, Paul Scherrer Institut (PSI) in Switzerland. A Ge (311) monochromator was chosen to obtain neutrons with the wavelength of 1.178 $\AA$ and a helium cryostat was used to achieve a base temperature of 2 K.
      
\textbf{M$\ddot{o}$ssbauer Measurements}. M$\ddot{o}$ssbauer spectra were collected on a polycrystalline sample of ScFeGe with a  Wissel spectrometer in the constant-acceleration mode over a range of temperatures between 8 and 295 K. A cobalt source in a rhodium matrix was used for the measurements, and $\alpha$-Fe was employed as a reference material for the isomer shift. The spectra were fit using a static Hamiltonian with a Lorentzian line shape. The fitting
procedure adjusts parameters of the hyperfine Hamiltonian listed in the Supplemental Material\cite{Supplement}, consisting of the isomer shift (IS), the quadrupole splitting (QS), the hyperfine field (HI), the electric field gradient (EFG) tensor asymmetry parameter ($\eta$), the angle between the hyperfine field and main axis of the EFG ($\Theta$), as well as the background count level, intensity of the absorption spectrum and the experimental line-width (W). This approach allows the calculation of the expected spectral shape for any
combination of magnetic dipole and electric quadrupole components contributing to the hyperfine Hamiltonian. It allows a seamless description of a pure quadrupole spectrum above the magnetic ordering temperature and a combined magnetic and quadrupole splitting below the critical point. This experiment was performed using the ``thin absorber'' technique, where the sample area density is less that 10 mg of Fe/cm$^2$ to guarantee a Lorentzian line shape of the spectral components. The best fit of the calculated spectrum to the experimental data was determined via the $\chi^2$ criterion.

\textbf{X-ray Absorption Spectroscopy Measurements}. Scandium, iron, and germanium \textit{K}-edge X-ray absorption near edge structure (XANES) spectroscopic measurements were performed at the J. Bennett Johnston, Sr., Center for Advanced Microstructures and Devices (CAMD) electron storage ring at Louisiana State University. Measurement details are very similar to those provided in the previous reports\cite{Haynes2019,Macheli2020}.  All measurements were made in transmission mode, and the data analysis was performed with Athena\cite{Ravel2005}.
   
\textbf{First-principles Calculations}. We performed first-principles calculations based on Density Functional Theory (DFT). Calculations of the structural, electronic, and magnetic properties of bulk ScFeGe were performed using the Vienna \textit{Ab Initio} Simulation Package (VASP)\cite{Kresse1993, Kresse1994, Kresse1996,KresseFurthm1996}. Ion cores were modeled with projector augmented wave (PAW) pseudopotentials\cite{Blochl1994}. The valence $3s$, $3p$, $3d$, and $4s$ states of scandium, the $3p$, $3d$, and $4s$ states of iron, and the $3d$, $4s$, and $4p$ states of germanium are treated explicitly. A plane-wave basis energy cutoff of 520 eV and a Gaussian smearing of 0.01 eV were found to yield
converged total energy and forces. For comparison purposes, we used the Perdew, Burke, and Ernzerhof (PBE) functional\cite{Kresse1999}, and we included van der Waals (vdW) interactions in our calculations. We also used the vdW density functional (vdW-DF), vdW-DF2, optPBE-vdW, optB88-vdW, and optB86b-vdW schemes, as well as DFT-D2 and DFT-D3\cite{Klimes2009,Klimes2011,Bucko2010,Grimme2010}. The spin-orbit coupling (SOC) was included in the self-consistent
calculations. All atoms, as well as the cell, were relaxed to force a cutoff of 10$^{-2}$ eV \AA$^{-1}$. The $k$-point sampling was based on a $\Gamma$-centered grid for all calculations. To optimize the primitive cells we used a $(10\times 10\times 10)$ grid and a $(7\times 7\times 7)$ grid for the supercells. Maximally-localized Wannier functions based on Sc-$d$, Fe-$d$, and Ge-$p$ orbitals were generated using the Wannier90 code\cite{Mostofi2014,Pizzi2019}. 

\begin{figure*}
\begin{center}
\includegraphics[width=6.0in]{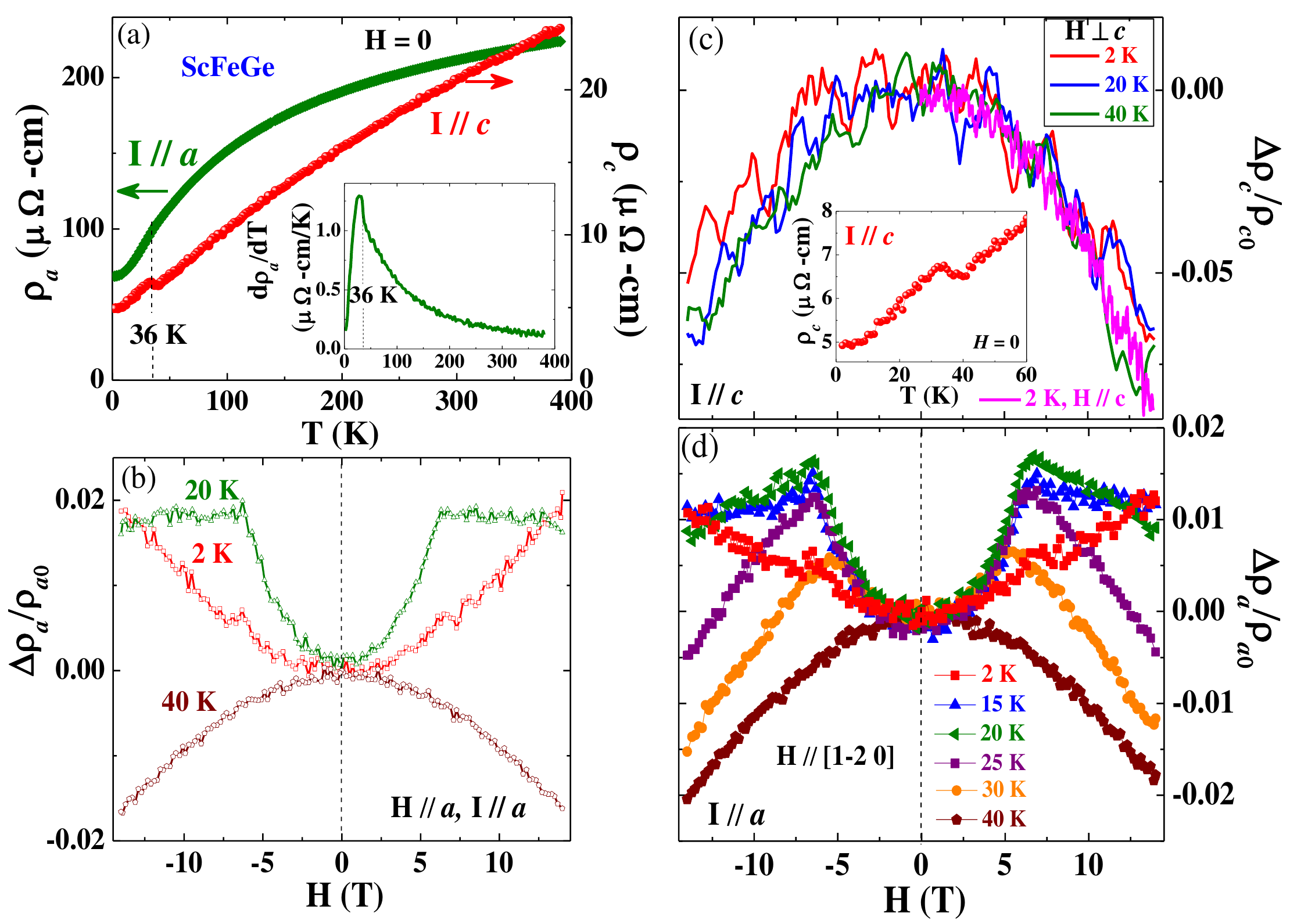}
\end{center}
\caption{\label{fig-MR}Charge transport. (a) Temperature, \textit{T}, dependence of the resistivity with the current, \textit{I}, parallel to the \textit{a}-axis, $\rho_a$, and current parallel to the $c$-axis, $\rho_c$, at zero field. Inset: $d\rho_a/dT$ vs. $T$. (b)  MR for current parallel to the $a$-axis, $\Delta\rho_{a}$/$\rho_{a}$(0) = ($\rho_{a}$(H)-$\rho_{a}$(0))/$\rho_{a}$(0), at the indicated temperatures. Data shown for \textit{H} parallel to the \textit{a}-axis. (c) Magnetoresistance (MR) for current parallel to the $c$-axis, $\Delta\rho_{c}$/$\rho_{c}$(0) = ($\rho_{c}$(H)-$\rho_{c}$(0))/$\rho_{c}$(0), at the indicated temperatures. Data are shown for field, $H$, parallel and perpendicular to the $c$-axis, as indicated in the figure. Inset: $\rho_c$ on an expanded scale at low $T$. (d) MR for current parallel to the $a$-axis, $\Delta\rho$/$\rho_0$ = ($\rho_{a}$(H)-$\rho_{a}$(0))/$\rho_{a}$(0), at the indicated temperatures. Data shown for \textit{H} applied along the [1 $\bar{2}$ 0].} 
\end{figure*}

\begin{figure}
\begin{center}
\includegraphics[width=3.0in]{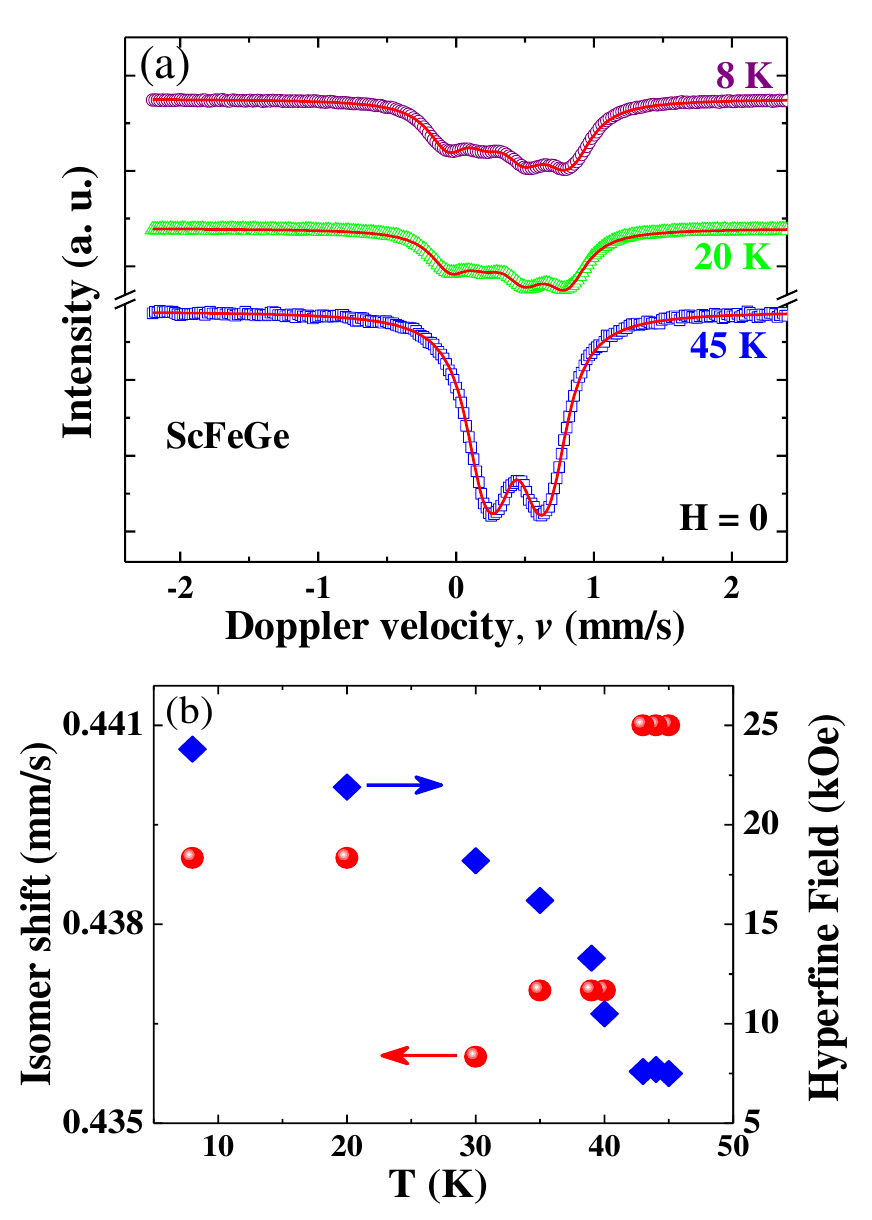}
\end{center}
\caption{\label{fig-MS} M$\ddot{o}$ssbauer spectroscopy. (a) M$\ddot{o}$ssbauer spectra of ScFeGe collected above and below the magnetic ordering at 36 K. (b)\textit{T} dependence of the isomer shift (left) and hyperfine field.}
\end{figure}

\begin{figure*}
\begin{center}
\includegraphics[width=6.0in]{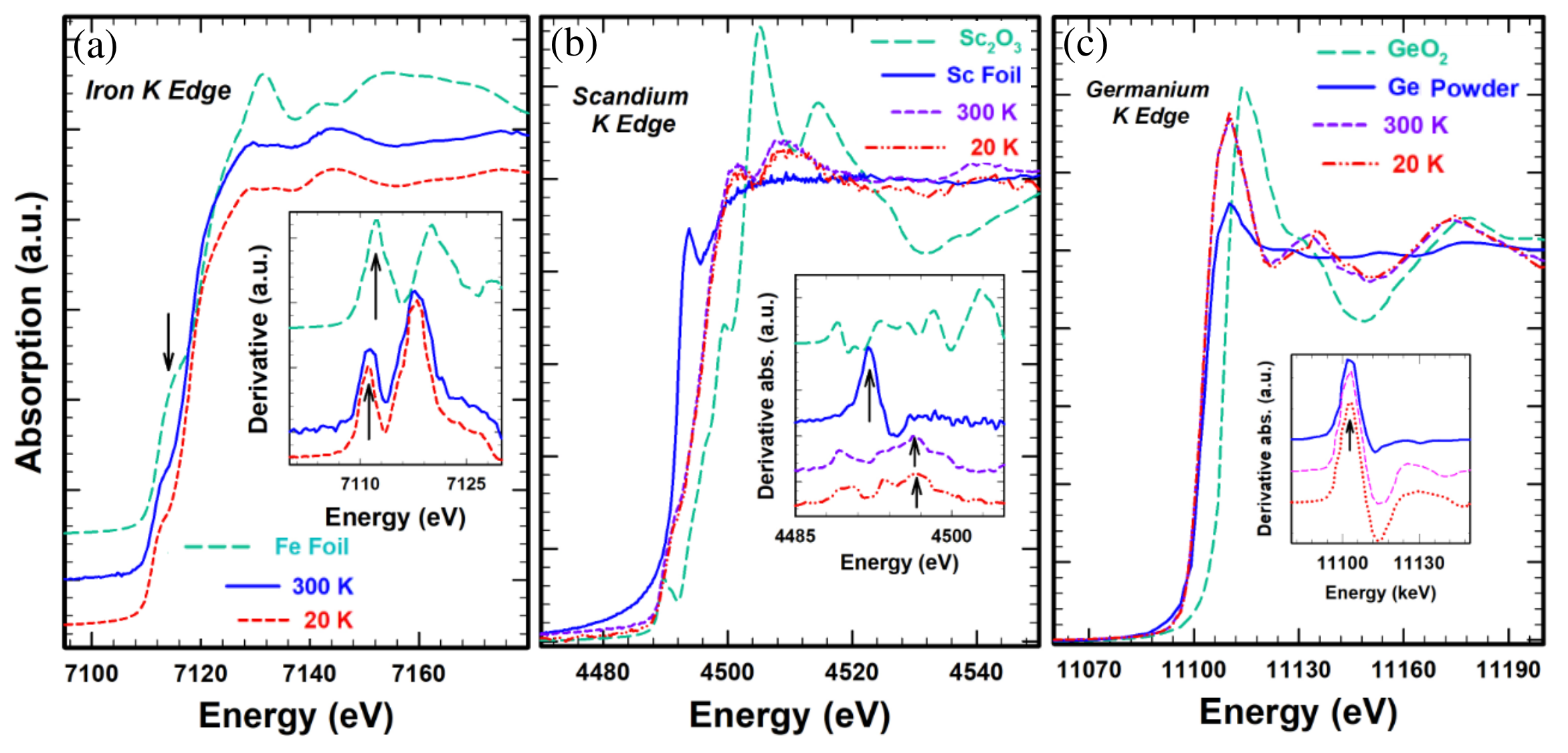}
\end{center}
\caption{\label{fig-XAS} X-ray absorption spectroscopy near edge structure (XANES). (a) Room temperature (RT) and low temperature 
(LT) (20 K) $K$-edge of Fe from ScFeGe (blue and red, respectively) and iron foil (Fe metal, green); inset shows the first-derivative edge positions. (b) RT and LT $K$-edge of Sc from ScFeGe (purple and red, respectively), scandium foil (Sc metal, blue), and in scandia (Sc$_2$O$_3$ , green); inset shows the first-derivative edge positions. (c) RT and LT K-edge of Ge from ScFeGe (purple and red, respectively), germanium powder (Ge, blue), and germania (GeO$_2$, green); inset shows the first-derivative edge positions.}
\end{figure*}

\section{Experimental Results} 
     
We begin the discussion of the magnetic properties of ScFeGe by presenting the temperature dependence of the magnetic susceptibility, $\chi$, of a single crystal for \textit{H}= 1 T, parallel and perpendicular to the \textit{c}-axis in Fig.~\ref{fig-chiT}(a). The sharp peak in $\chi$ at $T_{N}$ = 36 K, when $H$ is parallel to the $a$-axis, and the saturation of $\chi$ below this same temperature for $H$ $\parallel$ $c$-axis, indicate a magnetic transition at this temperature. A fit to the modified Curie-Weiss (CW) law, $\chi$ = $\chi_0 + C/(T-\Theta_W)$, where $C$ is the Curie constant, $\Theta_W$ is the Weiss temperature,
and $\chi_0$ the $T$- independent contributions to $\chi$, above 50 K yields $\Theta_W$ = -29.8(2) K for $H$ $\parallel$ \textit{c}-axis  and $\Theta_W=-10.49(7)$ K for $H\parallel a$-axis.  These indicate an antiferromagnetic interaction of the magnetic moments. The best fit resulted in $C = 0.666(2)$ emu K/mol with an effective magnetic moment $\mu_{eff} = 2.32(13) \mu_{B}$ for $H \parallel$ $c$-axis and $C = 0.606(1)$ emu K/mol, $\mu_{eff} = 2.21(8)\mu_{B}$ for $H \parallel$ $a$-axis. These data indicate a moderate anisotropy for all temperatures, with a cross over below 28 K. The transition temperature was found to vary between 36 and 38 K for our single crystal samples, while polycrystalline samples displayed a wider variation of $T_N$ up to 47 K with $T_N$ correlating well with the excess Ge content.
 
 The magnetic field dependence of the magnetization, $M$, of a single crystal sample is presented in Fig.~\ref{fig-chiT}(b) for fields up to 7 T.  As in $\chi(T)$, $M(H)$ displays a moderate anisotropy. Interestingly, a sharp metamagnetic (MM) transition  is observed at $H_{MM} = 6.7$ T for $T= 2$ K, when the field is oriented parallel to the \textit{a}-axis, but is absent when $H \parallel$ to the $c$-axis. $M(H)$ increases with field so that by 7 T,  $M$= 0.18 $\mu_B$, far below the value of the fluctuating moment determined from the Curie-Weiss fit, which indicates that a much larger magnetic field is required to saturate the magnetization. The MM transition broadens and shifts to lower fields with increasing temperature, as highlighted in Fig.~\ref{fig-chiT}(d) so that it appears at a field of 1.9 T for $T = 35$ K. Both $\chi(T)$ and $M(H)$ indicate an antiferromagnetic-like state with a small magnetic moment apparent even in a field of 7 T.

The specific heat at zero field is presented in Fig.~\ref{fig-chiT}(c), where $C_p$/T vs. $T^2$ is shown. A sharp peak
in $C_p$ at $T_N$= 36 K indicates that the magnetic transition discovered in $\chi(T)$ is intrinsic to ScFeGe. We highlight 
the low temperature, $T<17$ K, $C_p$/T in the inset of Fig.~\ref{fig-chiT}(c), where a linear fit (red line) reveals an intercept, $\gamma = 52.8(5)$ mJ/mole-$K^2$ and a linear coefficient, $\beta = 0.256(3)$ mJ/mole-$K^4$. The Debye temperature, $\Theta_D$, can be estimated using the relation $\Theta_D^3 = 12n\pi^4k_B/5\beta$, where $n$ is the number of atoms per formula unit, and $k_B$ is the Boltzmann constant yielding $\Theta_D = 196$ K. $\gamma$ is significantly larger than that found in other magnetic metals with similar ordering temperatures, such as MnSi ($T_N$ = 29 K and $\gamma = 32$ mJ/mole-$K^2$), indicating a larger spin-wave contributions to $C_p$ in ScFeGe.

The resistivity, $\rho$, of ScFeGe crystals with current oriented along the crystallographic $a$- ($\rho_a$) and $c$-axes 
($\rho_c$) was explored to characterize the anisotropy in the electronic properties, as well as the coupling between the magnetic and charge degrees of freedom. The temperature dependence of $\rho_{a}$ and $\rho_{c}$ at zero field are presented in Fig.~\ref{fig-MR}(a), where both exhibit metallic behavior with a residual resistivity ratio (RRR) of 4.2 for 
$\rho_{c}$ and 3.1 for $\rho_{a}$. $\rho_{c}$ is
found to be 13.9 times smaller than $\rho_{a}$ at low $T$, indicating a significant anisotropy in the electronic structure.
There is also a significant anomaly in $\rho_{c}$ at $T_N$
(see inset of Fig.~\ref{fig-MR}c), as cooling through $T_N$ results in a sharply increased resistivity. 
This suggests a significant 
reduction in the density of states at the Fermi level or even the opening of a partial gap in the electronic structure associated with the magnetic ordering. In contrast, $\rho_{a}$ displays only a peak in $d\rho_{a} / dT$ (Fig.~\ref{fig-MR}a) at $T_N$, without 
any significant discontinuity apparent. We note this difference in response to the magnetic ordering in the charge
conductivity along the $c$-axis and perpendicular to it indicates that the magnetic ordering affects mainly the 
charge carrier scattering and perhaps the electronic structure along the $c$-axis.  

\begin{figure}
\begin{center}
\includegraphics[width=3.0in]{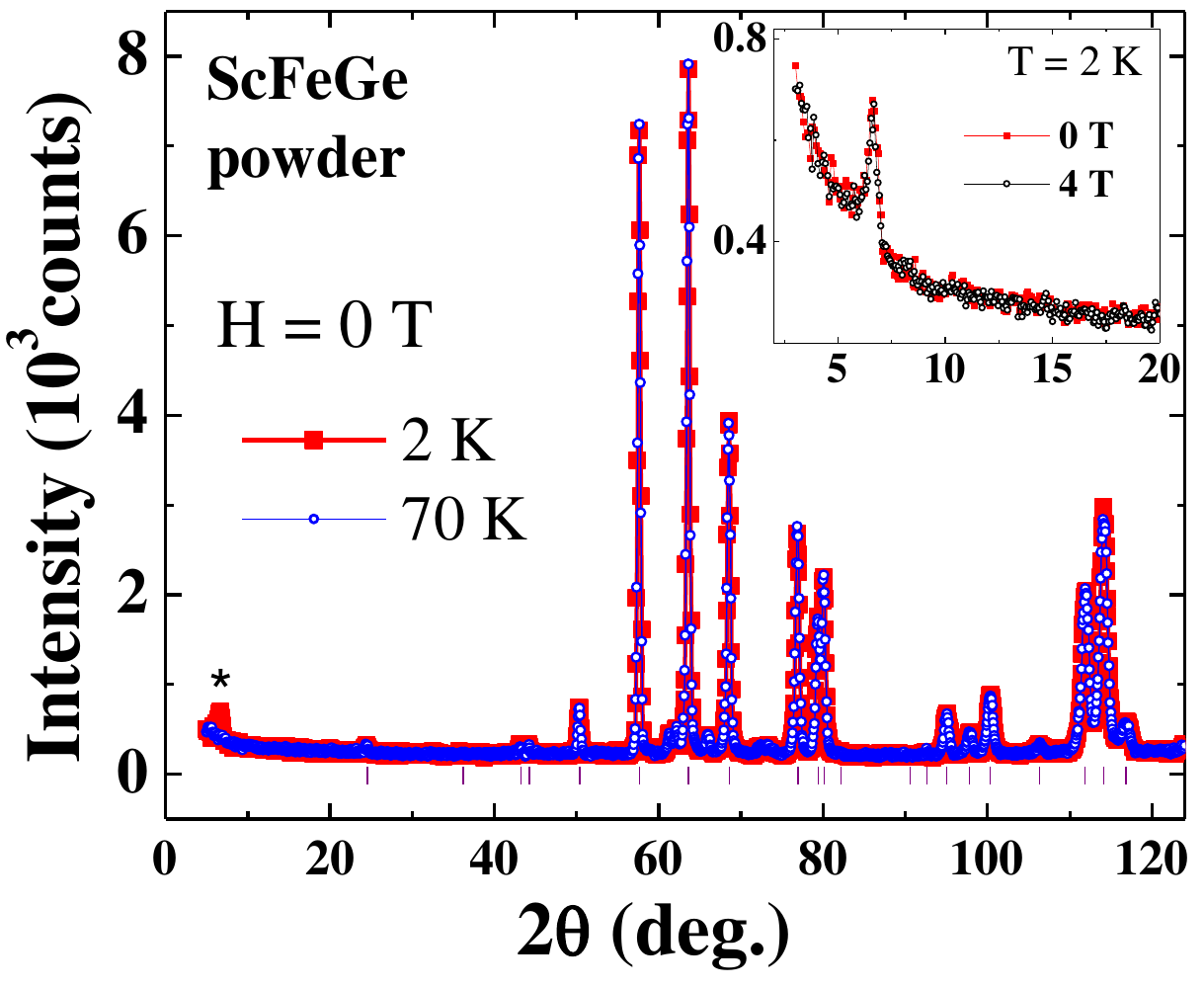}
\end{center}
\caption{\label{TDepNPD} Neutron powder diffraction data collected at zero field above (70 K) and below (2 K) $T_N$. Asterisk indicates the single magnetic diffraction peak. Inset shows the magentic peak of Neutron powder diffraction at 2 K in magnetic fields (\textit{H}) of 0 and 4 T. } 
\end{figure}
\begin{figure}
\begin{center}
\includegraphics[width=3.5in]{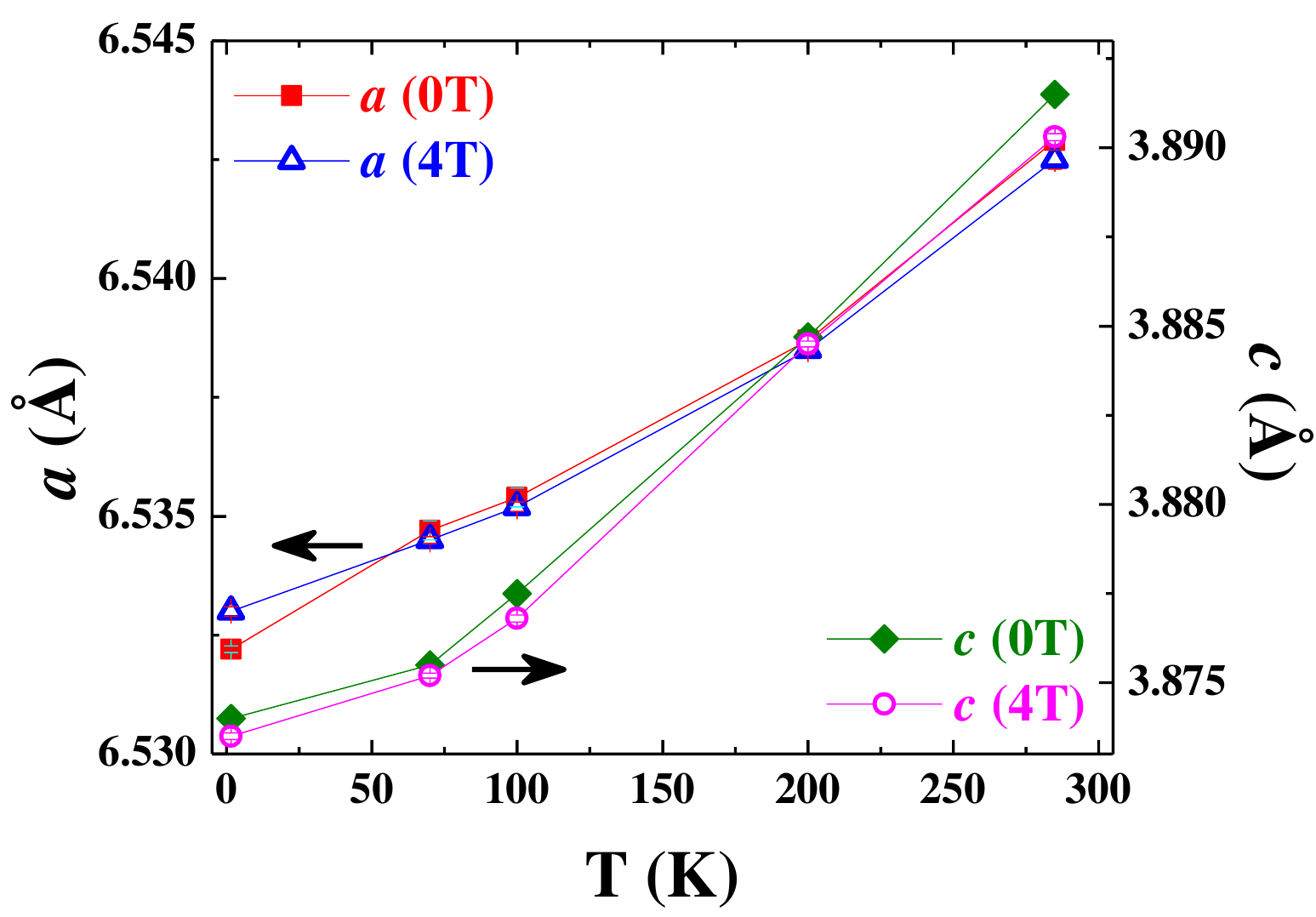}
\end{center}
\caption{\label{LatticeConst} Temperature, $T$, dependence of the lattice parameters \textit{a} and \textit{c} at magnetic fields, \textit{H}, of 0 and 4 T for ScFeGe. }
\end{figure}

The magnetic field dependence of $\rho_c$ for $H \le 14$ T
is presented in Fig.~\ref{fig-MR}(c) for $H$ parallel (longitudinal magnetoresistance, MR) and perpendicular (transverse MR) 
to the $c$-axis. We observe a moderately sized 
negative MR consistent with a fluctuation scattering mechanism that is nearly independent of the field direction. 
In contrast,  $\rho_a(H)$ (Fig.~\ref{fig-MR}b and d) displays much richer behavior. Here, $\rho_a(H)$ is shown with $H$ 
along the \textit{a}-axis (longitudinal MR) (Fig.~\ref{fig-MR}b) and along the [1 $\bar{2}$ 0] (transverse MR)(Fig.~\ref{fig-MR}d). These are similar with only a minor variation with field direction observed. This suggests that 
an interaction of the charge carriers with magnetic moments, rather than an orbital effect of the field, determines the MR. $\rho_a(H)$ is unusual in that there is a change in sign from a positive MR below $T_N$ to negative above $T_N$, in contrast to expectations for a simple fluctuation scattering mechanism, where a continuous suppression of scattering is common with cooling or increased $H$. The similarity of the MR for  $H > H_{MM}$ above and below $T_N$, along with the loss of the
negative contribution to $\rho_a(H)$ at low $T$, are supportive of a fluctuation scattering mechanism as the cause of the negative MR. The positive MR, which is only observed for $T< T_N$, is atypical of simple metallic magnets. That it is intimately tied to the magnetic structure is made clear both by its appearance below $T_N$ and the change in character of the MR at the MM transition. Interestingly, the discontinuity at $H_{MM}$  is lost upon cooling to 2 K. This is in contrast to the magnetization, which displays a trend toward a discontinuous transition at $H_{MM}$ at low $T$. The origin of the positive MR is not yet clear. 
              
We have also performed M$\ddot{o}$ssbauer measurements to further investigate the magnetic phase transition and to probe the character of the magnetism.  The M$\ddot{o}$ssbauer spectra recorded from a polycrystalline ScFeGe sample is shown in Fig.~\ref{fig-MS}(a). We observe a symmetric doublet at $H=0$ and 45 K that becomes asymmetric as the doublet splits in a hyperfine field when cooled below $T_N$. The shape of the spectrum at each of the three temperatures, where data was collected, is well represented by a fit of the model presented in the supplemental material~\cite{Supplement}. The best fit parameters include a hyperfine (HF) field, which increases from 8 to 25 kOe as $T$ is reduced from 45 to 8 K, as shown in Fig.~\ref{fig-MS}(b). This small value of the hyperfine field in the magnetically ordered state is usually indicative of a small ordered magnetic moment. For example, it is much smaller than that found in ScFe$_6$Ge$_6$ formed from the same
elements~\cite{Mazet2000,Samir2015}. This conclusion is at odds with the effective magnetic moment of $\sim 2.3 \mu_B$ found from fits to $\chi$ above $T_N$, suggesting a difference between the ordered and fluctuating magnetic moments, as is common in weak itinerant magnets. The isomer shift is plotted in Fig.~\ref{fig-MS}(b), where values similar to that found in
ScFe$_6$Ge$_6$~\cite{Mazet2000,Samir2015} are displayed. The isomer shift undergoes only small variation as it is cooled through $T_N$, suggesting only minor changes to either the crystal or electronic structures with magnetic ordering.

X-ray absorption near edge structure (XANES) spectroscopic measurements were performed to investigate the oxidation/electronic state of Sc, Fe, and Ge atoms in ScFeGe and to explore the possibility of differences between the sample at room temperature (RT; $T\sim 300$ K) and low temperature (LT; $T\sim 20$ K). In the case of the Fe $K$-edge (Fig.~\ref{fig-XAS}(a)), the near overlap of the XANES edge structure for ScFeGe and the metallic Fe standard indicates the Fe valence is close to zero in our compound.  However, as seen in the first-derivative plot in the inset, the small inflection (indicated by the arrows), which is due to 4s-3d mixing, is energetically slightly smaller (7111 eV in ScFeGe and 7112 eV in metallic Fe), indicating the possibility of a small reduction (corresponding to the gain of electrons) in the absorption of ScFeGe. This observation is consistent with the Bader charge analysis based on the electronic structure calculations, described below, wherein Fe gains 0.23 electron in ScFeGe. For the case of Sc, the $K$-edge in ScFeGe (Fig.~\ref{fig-XAS}(b)) is shifted with respect to the metal [Sc(0)] and the oxide [Sc(III)]. For Sc$_2$O$_3$ there is a strong pre-edge feature near $\sim 4490$ eV, which can better been seen in the first-derivative plot (inset). This pre-edge structure is due to a crystal-field splitting of the Sc 3d orbitals.  The Sc-edge for ScFeGe also shows a small, less intense inflection.  From the maximum in the first-derivative plots, wherein the Sc(0) peaks at $\sim 4492$ eV and Sc(3) at $\sim 4503$ eV, a simple linear interpolation indicates that the Sc in our compound (with a peak at $\sim 4496$ eV) has an oxidation of $\sim +1.2$.  This observation is also consistent with the Bader charge analysis based on the calculations described below, wherein Sc loses 1.43 electrons. Finally, in the case of Ge (Fig.~\ref{fig-XAS}(c)), the ScFeGe $K$-edge structure appears different from the Ge(0) powder, despite the similarity in the edge energy position (i.e. inflection point) at 11,103 eV.  The oxidation of Ge is more difficult to determine from the edge energy position because of the covalent bonding. For example, although the K-edge is approximately  $\sim 6$ eV less than that of Ge(4) in GeO$_2$, other model compounds, such as GeS, have a $K$-edge at 11,103 eV, much closer to our value. Because of this, it is difficult to determine the oxidation of Ge in our compound. The Bader charge analysis indicates that Ge gains 1.2 electrons in this compound.
As seen in Figs.~\ref{fig-XAS}(a-c), the Fe, Sc, and Ge $K$-edges, and absorption structure of ScFeGe, are very similar at both RT and LT, indicating no apparent oxidation changes. Thus, we are not able to resolve any significant changes to the electronic structure above and below the transition temperature (36 K) in our XANES measurements.

Neutron powder diffraction measurements were performed over a wide wavevector, $Q$, range to determine the magnetic structure that emerges in ScFeGe below $T_N$ (Fig.~\ref{TDepNPD}). Upon cooling ScFeGe to 2 K, a new satellite peak appears in the diffraction pattern at $2\theta$= 6.5$^\circ$ (indicated by an astrisk in Fig.~\ref{TDepNPD} and in Fig.~\ref{NPD}b) that was absent at 285 K. No other peaks appear above background at higher $Q$, suggesting that this additional scattering peak is magnetic in origin. In addition, we
observe no change to the scattering intensity at the positions of the nuclear Bragg peaks, ruling out a substantial ferromagnetic component to the magnetic structure (see Fig.~\ref{TDepNPD}). The small angle at which the new scattering peak appears suggests a helical or spin density wave ordering with a pitch length or wavelength of $\sim$ 20 \AA. Application of a 4-T magnetic field had no significant effect on either the angle or the intensity of the magnetic scattering peak discovered (inset of Fig.~\ref{TDepNPD}), suggesting a much higher field is necessary to field polarize this magnetic state, in agreement with $M(H)$ measurements (Fig.~\ref{fig-chiT}(b)). The temperature dependence of the lattice parameters in zero and 4-T fields, as determined from the powder neutron diffraction data, is shown in Fig.~\ref{LatticeConst}. No significant change in either $a$ or $c$ is observed, either as a result of the magnetic ordering at $T_N$, or the application of the magnetic field.

\begin{figure*}
\begin{center}
\includegraphics[width=6.5in]{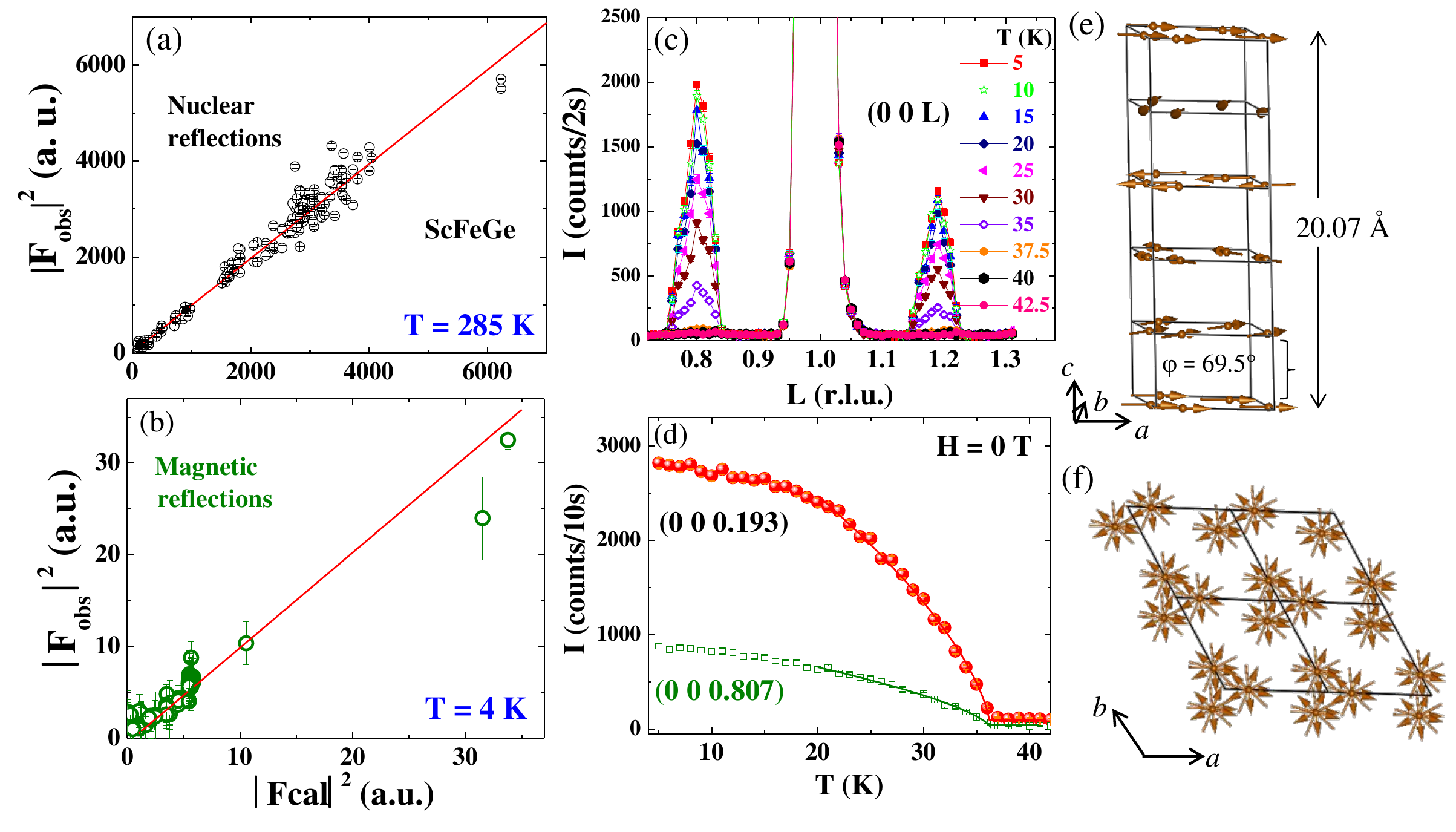}
\end{center}
\caption{\label{fig-MagneticStru} Single-crystal neutron diffraction and magnetic structure. (a) Comparison of the calculated and observed values of the squared structural factors at \textit{T} = 285 K. The model for the hexagonal crystal structure with the $P\bar{6}2m$ space group and structural parameters determined from the PXRD at room temperature, and listed in Table S1 in the supplemental material~\cite{Supplement}, were used to calculate the expected intensity. (b) Comparison of the calculated and observed values of the squared structural factors for the incommensurate phase established at T = 4.0 K at zero field. A helical magnetic structural model was employed in the calculation. (c) L-scans near the (001) Bragg peak at several representative temperatures. (d) \textit{T} dependence of the peak intensity of (0 0 0.193) and (0 0 0.807) incommensurate magnetic peaks, indicating magnetic ordering T$_{N}\sim$ 36 K. The solid line is a fit of a standard critical behavior model to the intensity (see text). (e) Schematic demonstrating the incommensurate helical magnetic structure at zero field. (f) View of the magnetic structure along the $c$-axis. } 
\end{figure*}

Because of the limited information provided by powder diffraction, which displayed only a single magnetic scattering peak, single-crystal neutron diffraction was carried out to establish more clearly the magnetic structure. To this end, a set of 105 reflections was collected at both 4 and 285 K as input for the refinement of the crystal and magnetic structures. The quality of the refinement of the crystal structure can be seen in Fig.~\ref{fig-MagneticStru}(a). The refinement of the magnetic structure
indicates an incommensurate helical structure along the $c$-axis with wave vector $\bm k$ = (0 0 $\delta$) with $\delta = 0.193$. A magnetic moment of $\mu_S$ = 0.53 $\mu_B$ per formula unit (f.u.), confined to the $ab$-plane was determined. The quality of the refinement is demonstrated in Fig.~\ref{fig-MagneticStru}(b), and schematics of the magnetic structure are included in Fig.~\ref{fig-MagneticStru}(e) and (f). Just as was true of the powder neutron diffraction data, we observe no discernible changes to the nuclear Bragg scattering peaks that could be associated with a ferromagnetic component to the magnetic structure (see e.g. Fig.~\ref{TdepNuclCrys})\cite{Supplement}. In addition, a number of scans along the (0 0 L) direction were performed as a function of temperature to characterize the changes to the magnetic state with warming through $T_N$. Fig.~\ref{fig-MagneticStru}(c) presents these data in the vicinity of the (0 0 1) nuclear Bragg peak, where the incommensurate peaks at (0 0 $1$-$\delta$) and ((0 0 $1$+$\delta$) are reduced with temperature, until they are indistinguishable from the background above $T_N$. No change in $\delta$ was observed with warming.

The temperature dependence of the (0 0 $\delta$) and (0 0 $1$-$\delta$) satellite peaks is presented in Fig.~\ref{fig-MagneticStru}(d) along with fits of these data to the the standard form $I = I_0 + I_M(1-T/T_N)^{2\beta}$ as described in Ref.(47). Here, $I_0$ represents the temperature independent contribution from backgrounds, $I_M$ is the magnetic intensity at saturation, and $\beta$ is the critical exponent. The solid lines represent the best fit of this form to the data with a value for the critical exponent $\beta$ = 0.32(1) for both the (0 0 0.193) and (0 0 0.807) peaks at zero field. This value of $\beta$ lies between that expected for the three-dimensional Heisenberg model ($\beta$ = 0.38)~\cite{Ghosh1998} and the two-dimensional XY model ($\beta$ = 0.23)\cite{Nielsen1993}.

\section{Results of First-Principles Calculations} 

Electronic structure calculations were carried out to lend insight into the mechanisms for the electronic and magnetic properties presented in the previous section of this paper. The primitive cell used in our calculations was that of ScFeGe as described by the hexagonal lattice illustrated in Fig.~\ref{fig-crystru}(a). The optimized structural parameters ($a$ and $c$) and the ground state energy per atom ($E_0$) of ScFeGe that were calculated for several different spin orientations and exchange-correlation (xc) functionals are listed in Table~\ref{Tab1}. For the PBE calculations, we note that the primitive cell is energetically more stable by $\sim30$ meV/atom, when the magnetic moment is included. In our calculations, we do not observe any magnetic anisotropy. Therefore, to compare the differences between each xc functional, we choose to set up the spin orientation along the $a$-axis. In comparison with the experimental values of bulk ScFeGe, the PBE results along the $a$- and $c$-axis are underestimated by $0.7\%$ (6.485 \AA ) and overestimated by $2.4\%$ (3.967 \AA), respectively. When the van der Waals (vdW) corrections are included, the relative differences along the $a$- and $c$-axes increase by $3.33\%$ (6.750 \AA) and $5.29\%$ (4.079 \AA), respectively. In the case of DFT-D3, the $a$-axis is compressed by around $1.9\%$ (6.409 \AA), while the inclusion of the long-range vdW interactions expand the $c$-axis by only $0.6\%$ (3.897 \AA). Taking the measured lattice constants from ScFeGe as a reference suggests that the semilocal PBE functional is the best option to calculate the physical properties of ScFeGe. The calculated magnetic moments, $m_{spin}$, for ScFeGe are also summarized in Table~\ref{Tab1}. We notice that all calculated magnetic moments deviate substantially from the experimental values for the fluctuating effective magnetic moment at $T>T_N$ ($\sim$ 2.3 $\mu_B$/f.u.) and the ordered magnetic moment determined from refinements of the neutron diffraction data ($\mu_S$ = 0.53 $\mu_B$/f.u.).

\begin{table}[t!]
\caption{Results of electronic structure calculations for different spin orientations and exchange-correlation functionals. Lattice parameters, $a$ and $c$ (\AA) and ground state energy per atom, $E_0$ (eV/atom) of ScFeGe primitive cells and supercells, obtained by including spin-orbit coupling and magnetic moment, $m_{spin}$ ($\mu_{B}$/Fe atom). These theoretical data are compared with our experimental values measured at 2 K.}
\centering
\scalebox{0.92}{
\begin{tabular}{l l l l l l}
\hline
\hline
Spin orientation & Functional & $a$ & $c$ & $m_{spin}$ & $E_0$ \\
                            
\hline
Non-magnetic      & PBE         & 6.719 & 3.535 & 0.00 & -6.876 \\
         &             &       &       &      &        \\
$c$-axis & PBE         & 6.485 & 3.968 & 1.86 & -6.906 \\
         &             &       &       &      &        \\
$a$-axis & PBE         & 6.485 & 3.967 & 1.84 & -6.906 \\
         & vdW-DF      & 6.743 & 4.057 & 2.18 &        \\
         & vdW-DF2     & 6.750 & 4.079 & 1.94 &        \\
         & optPBE-vdW  & 6.629 & 4.032 & 1.97 &        \\
         & optB88-vdW  & 6.539 & 3.990 & 1.81 &        \\
         & optB86b-vdW & 6.542 & 3.991 & 1.81 &        \\
         & DFT-D2      & 6.401 & 3.972 & 1.78 &        \\
         & DFT-D3      & 6.409 & 3.897 & 1.52 &        \\
         &             &       &       &      &        \\
Helical-72$^{\circ}$   & PBE   & 6.469 & 3.984 & 1.96 & -6.908 \\
Helical-90$^{\circ}$   & PBE   & 6.461 & 3.998 & 2.00 & -6.909 \\
         &             &       &       &      &        \\
Expt. (NPD at 2 K)     &       & 6.532 & 3.874 & 0.53 $\mu_{B}$/f.u.  &    \\
                       &       &       &       &  & \\
\hline
\hline
\end{tabular}
}
\label{Tab1}
\end{table}    

\begin{figure}
\begin{center}
\includegraphics[width=3.5in]{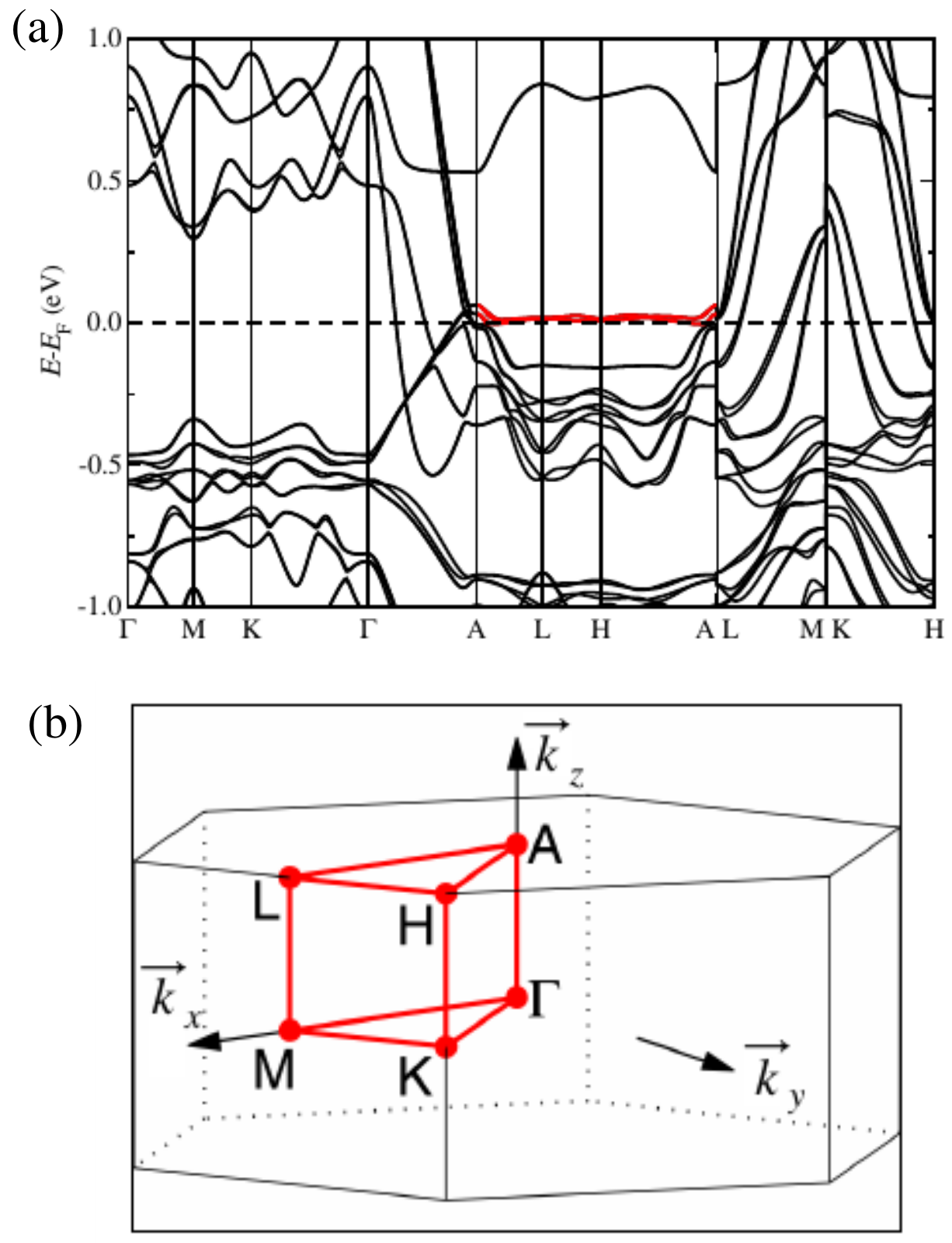}
\end{center}
\caption{\label{fig-BandStru} Band Structure. (a) Calculated electronic band structure of nonmagnetic ScFeGe for energies, $E$ within 1 eV of the Fermi energy, $E_F$. Spin-orbit coupling is included. $E_F$ is at zero energy. Highlighted in red are the nearly degenerate flat bands near $E_F$. (b) The first Brillouin zone of a hexagonal lattice.~\cite{Setyawan2010}  } 
\end{figure}

\begin{figure*}
\begin{center}
\includegraphics[width=6.0in]{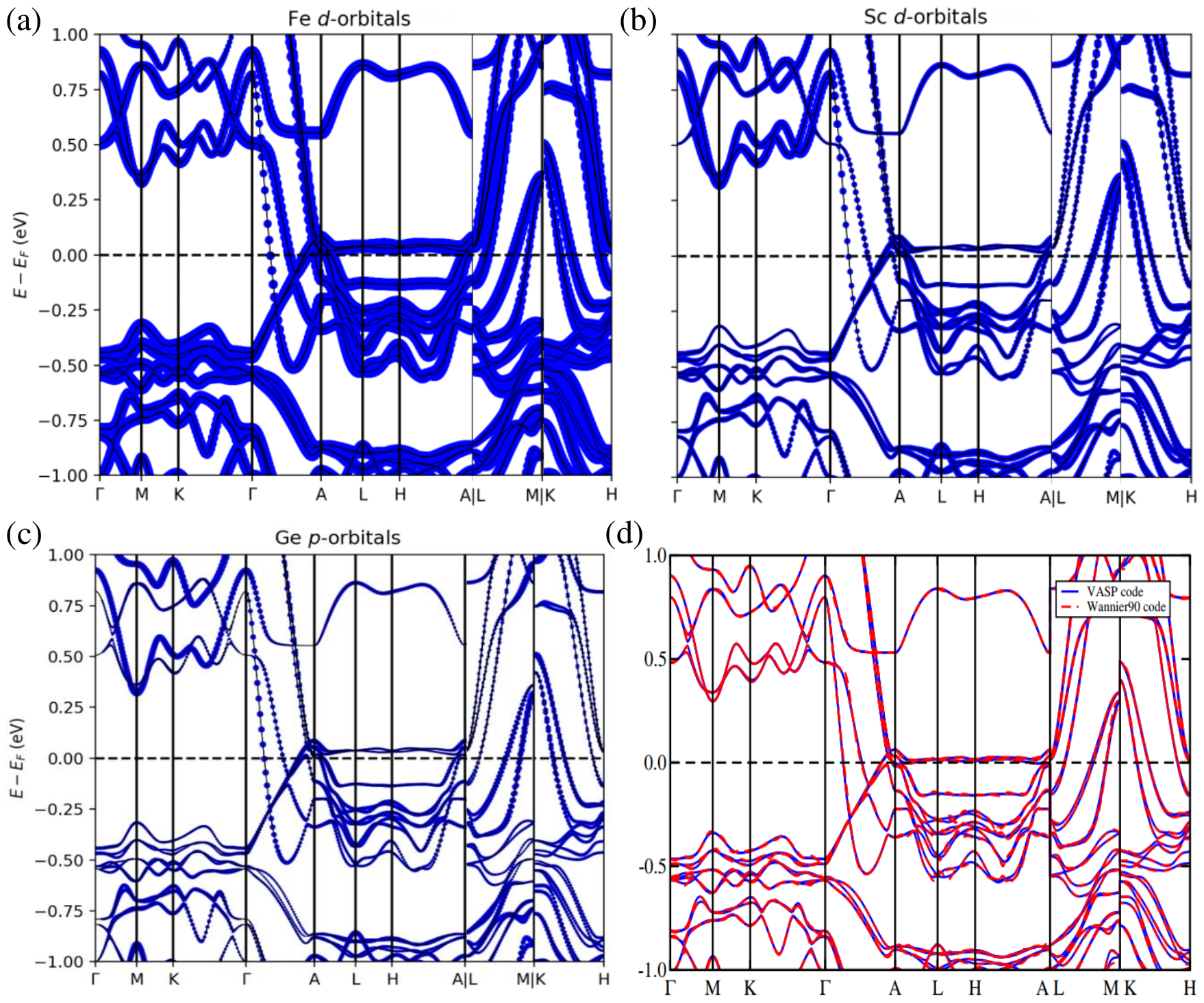}
\end{center}
\caption{\label{Fig-S1} Calculated electronic band structure of nonmagnetic ScFeGe showing orbitals of (a) Fe, (b) Sc, and (c) Ge  in which \textit{d}-type orbitals of the Fe atoms are the dominant feature for these bands. The size of the filled circles are proportional to the weight projected onto the orbitals. (d) Calculated electronic band structure of nonmagnetic ScFeGe, using first-principles calculations (blue line) and Wannier functions (red dashed line), where Kohn-Sham states are projected onto Fe-$d$, Sc-$d$, and Ge-$p$ localized orbitals. The spin-orbit coupling is included. The Fermi level is at zero energy. } 
\end{figure*}

The electronic band structure of nonmagnetic ScFeGe was calculated employing the PBE functional, as shown in Fig.~\ref{fig-BandStru}(a) for energies up to 1 eV above and below Fermi energy. There are several nearly degenerate flat bands (highlighted in red) located just above the Fermi level along the high-symmetry $\bm k$-path of A-L-H-A. This contour delimits the upper edge of the first Brillouin zone (BZ), as shown in Fig.~\ref{fig-BandStru}(b)\cite{Setyawan2010}. In addition, several valence and conduction bands cross the Fermi level mainly along $\Gamma$-A, L-M, and K-H lines, \textit{i.e.}, along the $k_z$-axis. The orbital-character analysis was performed to establish the character of the bands closest to the Fermi level. As shown in Fig.~\ref{Fig-S1}, the $d$ orbitals of Fe, and in particular the $d_{z^2}$ state, are dominant in the vicinity of the Fermi level. In addition, the electronic bands located between 0.5 and 1 eV above the Fermi level derive mainly from Sc-$d$ and Ge-$p$ orbitals. A Bader charge analysis was performed that showed that Fe gains 0.23 electron, Ge gains 1.20 electrons, and Sc loses 1.43 electrons. These are in general agreement with the XANES results, but we note that the XANES result for Ge was inconclusive. 

\begin{figure*}
\begin{center}
\includegraphics[width=6.0in]{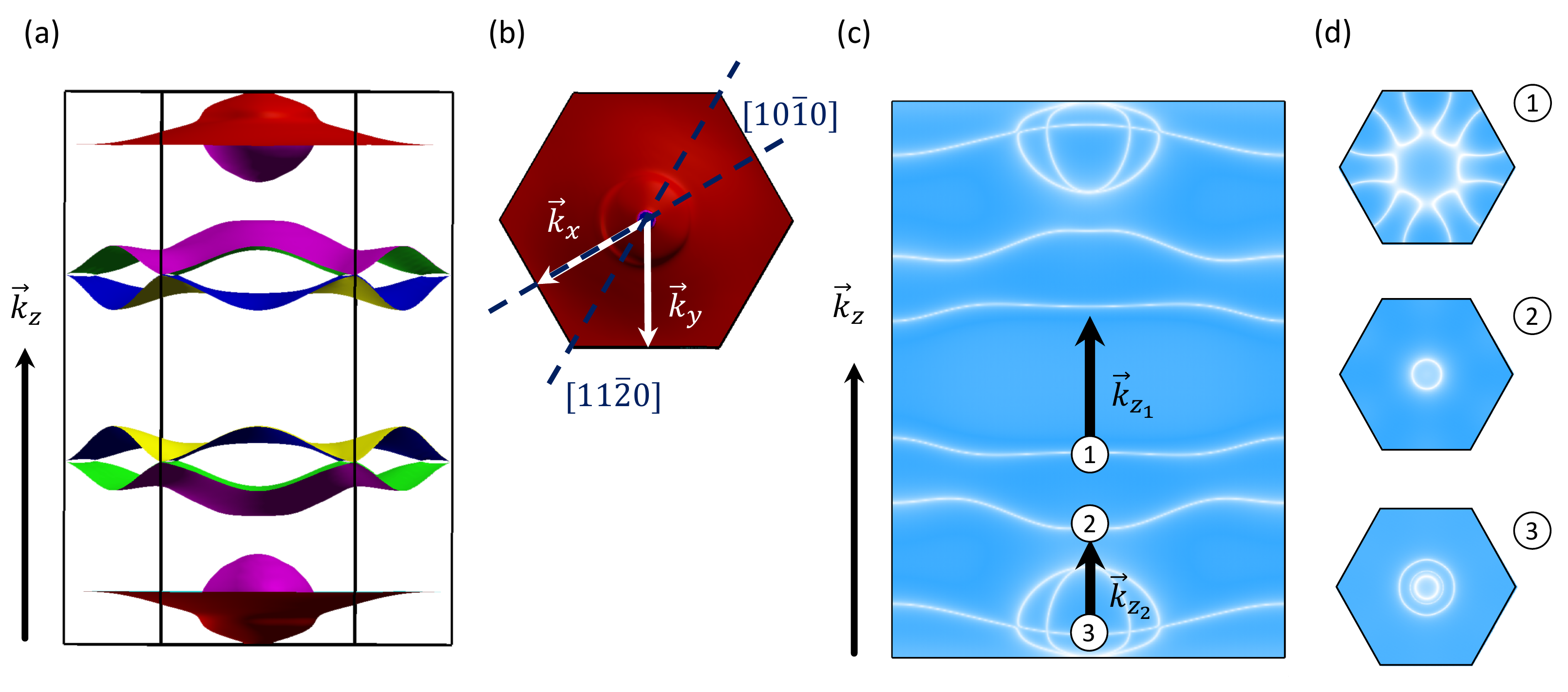}
\end{center}
\caption{\label{fig-FermiSur} Fermi surface. (a) Side and (b) top views of the calculated three-dimensional Fermi surface (FS) sheets in the first Brillouin zone for nonmagnetic bulk ScFeGe. (c) Calculated bulk FSs on a cleaved [1 0 $\overline{1}$ 0] surface. The coordinates of $k_{z_1}$ and $k_{z_2}$ are (0 0 0.262) and (0 0 0.191), respectively. (d) Bulk FSs at $k_z = -0.136$, $-0.265$, and $-0.456$ and labeled as \textcircled{1}, \textcircled{2}, and \textcircled{3}, respectively. The spin-orbit coupling is included.} 
\end{figure*}

The  nearly degenerate flat bands close to the Fermi level suggest an electron-driven instability. One way to explore this possibility is to investigate the three-dimensional (3D) Fermi surface sheets (FSs). These were determined from the symmetrical Wannier tight-binding model to project the Kohn-Sham states onto Fe-$d$, Sc-$d$, and Ge-$p$ localized orbitals. The resulting electronic band structure is compared to that obtained from first-principles calculations in Fig.~\ref{Fig-S1}(d), where it can be seen that the interpolated bands reproduce the electronic structure of ScFeGe in the range between -1 and 1 eV very well. The resulting Fermi surfaces are depicted in Fig.~\ref{fig-FermiSur}(a-b), showing nearly flat sheets perpendicular to $k_z$-axis, as suggested by the band structure of Fig.~\ref{fig-BandStru}(a), along with two ellipsoidal Fermi surface sheets centered at the top and bottom of the BZ. For better visibility, we project the FS onto the [1 0 $\overline{1}$ 0] plane, as shown in Fig.~\ref{fig-FermiSur}(c). Along the $\Gamma$-A line, we can extract two wavevectors, $\bm k_{z_1}=$(0 0 0.262) and $\bm k_{z_2}=$(0 0 0.191), which connect the three flattest Fermi surface sheets. The magnitude of these vectors, particularly $\bm k_{z_2}$, is similar to the helical wavevector obtained from neutron diffraction.  $\bm k_{z_1}$ and $\bm k_{z_2}$ can be associated with turn angles of 94.32$^{\circ}$ and 68.76$^{\circ}$ between moments in adjacent planes. Fig.~\ref{fig-FermiSur}(d) shows three cuts through the FSs perpendicular to $\bm k_z$ at $\bm k_z = -0.136$, $-0.265$, and $-0.456$. It is clear that first-principles calculations predict a highly nested Fermi surface with nesting wavevectors similar to the incommensurate wavevector found in neutron diffraction, although our calculations overestimate the size of the moments in the magnetically ordered state. This, along with the anomaly observed in the resistivity for currents along the $c$-axis, strongly suggests a Fermi surface nesting instability as the mechanism driving the magnetic phase transition discovered in ScFeGe.

\begin{figure}
\begin{center}
\includegraphics[width=3.5in]{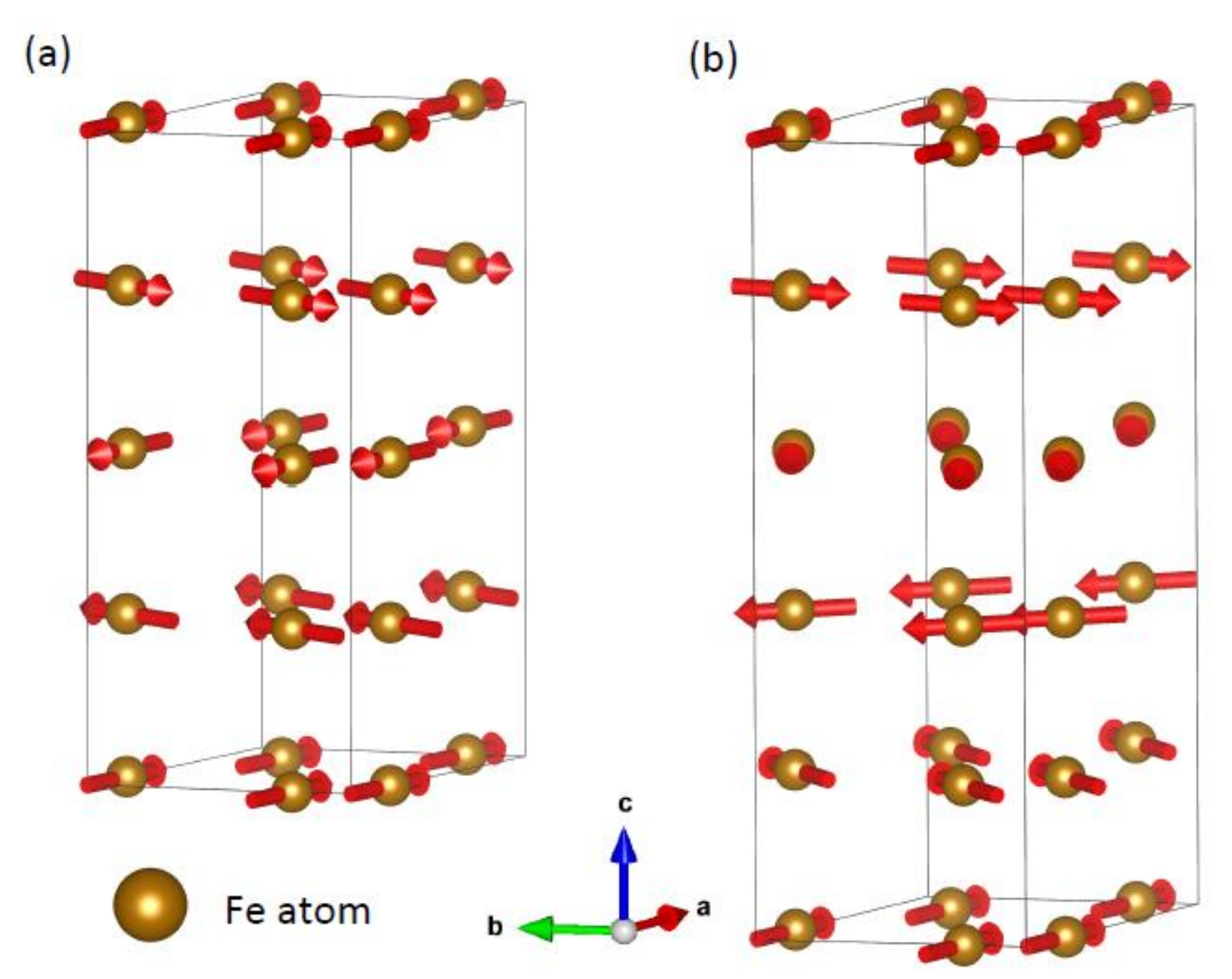}
\end{center}
\caption{\label{SpinStruTheo} Schematics of the supercells used in the (a) helical 90$^{\circ}$ (1$\times$ 1$\times$ 4) supercell, and (b) helical-72$^{\circ}$ (1$\times$ 1$\times$5) supercell calculations. For clarity, only the iron atoms are shown with the spin orientations represented by red arrows.} 
\end{figure}

To determine the stability of helical magnetic order at both $\bm k_{z_1}$ and $\bm k_{z_2}$, we have considered $(1\times 1\times 4)$ (90$^\circ$ moment rotation between adjacent planes) and $(1\times 1\times 5)$  (72$^\circ$ rotation) supercells, as shown in Figs.~\ref{SpinStruTheo} (a) and (b), respectively. These are commensurate structures that we use to approximate the predicted incommensurate ordering associated with the Fermi surface nesting to allow calculations to be performed. The optimized structural parameters for these calculations are included in Table \ref{Tab1}. Both supercell structures behave in a similar fashion in that the $a$-axis decreases ($\sim$0.25\%),
while the $c$-axis increases ($\sim$0.78\%), when compared with the calculated primitive cell of ScFeGe. This indicates that the non-collinearity of the spin orientation strongly influences the ScFeGe structure. Energetically, these two structures are equivalent to within our computational accuracy at -6.909 eV/atom, \textit{i.e.}, at zero temperature the 90$^{\circ}$- and 72$^{\circ}$-helical structures are nearly degenerate. However, for a finite transition temperature, the free energy, including the entropy, must be considered to determine which of these structures may be more stable. If we consider a Ginzburg-Landau expansion in powers of the
Fourier components of the magnetization, $m_{\bm k}$, the implication of the \textit{ab initio} calculations are that the coefficients of the $|m_{\bm k}|^2$ terms for both $\bm k_z=$(0 0 0.25) and $\bm k_{z}=$(0 0 0.2) change sign at approximately the same temperature. Since these leading instabilities compete for the same magnetic moment, their interaction is repulsive. If we assume that the energy landscape in proximity to the nesting vectors is not too steep,
then the result of this calculation is that there are closely competing helical magnetic states in ScFeGe. Given that only a single peak is observed in the neutron diffraction, the repulsive interaction between the two orders is likely sufficiently strong to prevent coexistence of both types~\cite{Liu1973,Kosterlitz1976}. In that case, the actual
transition temperature will be renormalized by the thermal fluctuations of the respective order parameters. In mode-mode coupling theory\cite{Murata1972,Lonzarich1985}, the thermal fluctuations are determined by the values of the coefficients of the quartic terms in the expansion and by the phase space available for fluctuations in the vicinity of  $\bm k_{z_1}$ and $\bm k_{z_2}$. The information on these coefficients is not contained in the \textit{ab initio} results, and one may plausibly argue that the lower symmetry 72$^{\circ}$ results in a sufficiently large phase space for fluctuations to bring the
transition temperature below that of the 90$^{\circ}$ order.

\section{Discussion and Conclusions} 

   In summary, we have discovered magnetic ordering in NCS hexagonal ScFeGe at $T_N=36$ K apparent in the magnetic susceptibility, magnetization, specific heat, electrical transport, M$\ddot{o}$ssbauer spectroscopy, and neutron diffraction measurements. In addition, a metamagnetic transition is found (at 6.7 T for $T=2$ K), when the field is oriented within the hexagonal \textit{ab} plane, but absent for $H$ parallel to the $c$-axis. Our neutron diffraction measurements reveal an incommensurate helimagnetically ordered state below $T_N$ with a wavevector \textbf{\textit{k}} = (0 0 0.193) and a magnetic moment of $\mu_S$ = 0.53 $\mu_B$/f.u. aligned within the $ab$-plane. The size of $\mu_S$ is in reasonable agreement with the small hyperfine field observed in our M$\ddot{o}$ssbauer measurement, but inconsistent with the effective magnetic moment determined from the Curie constant ($\mu_{eff} \sim 2.3 \mu_B/$ Fe atom).  The ratio of these moments, $RW = \mu_{eff} / \mu_S$, is known as the Rhodes-Wohlfarth parameter and is used to indicate the degree of itinerancy of the magnetic state\cite{Rhodes1963}. Here, $RW = 4.3$ is comparable to that observed in famous itinerant magnets ZrZn$_2$\cite{Chatterji2006,Pickart1964,Yelland2005}, MnSi\cite{Chatterji2006,Pffeiderer2001}, and Fe$_{1-x}$Co$_x$Si\cite{Kawarazaki1976}, indicating that ScFeGe belongs to the class of weak itinerant magnets. This is important, as such systems are typically sensitive to chemical substitution and pressure. As a result, many of these systems display quantum criticality with the possibility of unconventional superconductivity or other exotic phases. The tendency of our polycrystalline samples with an increased Ge content to display higher $T_N$ is consistent with this expectation.

The strong connection of the magnetism with the charge carriers is reflected in the MR as well. The MR displays unusual field and temperature dependencies for $T<T_N$ including the appearance of a positive MR below $T_N$ and at fields $H<H_{MM}$ for $H$ oriented perpendicular to the $c$-axis. The implication is that these fields cause an increase in scattering for the charge carriers and suggests a complex variation of the magnetic structure with field. In addition to the experimental data, our electronic structure calculations indicate an itinerant mechanism for the magnetic ordering in the form of a strongly nested Fermi surface with one nesting vector consistent with the neutron diffraction data. 

The helimagnetic ordering with a wavevector along the $c$-axis in a hexagonal system draws natural comparisons to the heavy rare earth elemental metals Tb, Dy, and Ho. These materials display incommensurate helimagnetic states over a finite temperature range below the critical temperature where there is a strong temperature dependence to the helimagnetic pitch length\cite{Chattopadhyay1994}. The ground states are ferromagnetic, as $H_{MM}$ is reduced to zero field with cooling, or there is a transition to a commensurate conical phase\cite{Dobrich2010,Tapia2017}. In addition, nesting conditions along the $c$-axis have been discovered in electronic structure calculations suggesting a strong connection of the magnetism to the charge degrees of freedom\cite{Evenson1969}. These materials have drawn more recent attention because of the complex magnetic structures that form when exposed to magnetic fields oriented perpendicular to the $c$-axis\cite{Chattopadhyay1994}. Here, fan, helifan, and spin-slip states have been discovered with abrupt transitions in magnetic fields\cite{Jensen1990,Chattopadhyay1994}.  There are also significant differences between these heavy rare earth metals and ScFeGe, including the character of the magnetism, as $f$-electron materials are thought of as having localized magnetic moments that interact via the conducting charge carriers. ScFeGe has a small ordered magnetic moment, so that a local magnetic moment model is not appropriate, and we have demonstrated that the helimagnetic pitch is not strongly temperature or field dependent. Despite these differences, the experience with the heavy rare earth elemental metals suggests complex magnetic structures may exist at finite field in ScFeGe. Thus, how the magnetic state of ScFeGe approaches the field polarized state at higher magnetic fields is an interesting unanswered question. The path to full polarization may be substantially different in ScFeGe than the classic cases discussed for the rare earth elemental metals based upon the weak itinerant character of its magnetism.

Although the motivation for this work was to explore a NCS magnet where the DMI is expected to play a role, it appears that the DMI may play only a minor role in the mechanism for helimagnetism in ScFeGe. In addition, helimagnetism is often described to occur due to competing interactions where there are exchange interactions of opposing sign and similar magnitudes. In this case, the expectation is that magnetic fields cause a distortion of the helical state causing regions where the magnetic moments are mostly along the field to grow at the expense of regions where the moments oppose the applied field\cite{Udvardi2006,Glasbrenner2014}. For materials that are well described by this model, a MM transition is found to occur to a magnetic fan state with $H_{MM} \sim 0.5 H_{sat}$, where $H_{sat}$ is the field necessary to saturate the magnetization\cite{Herpin1961,Nagamiya1962}. Instead, for ScFeGe, we find that the pitch of the helical state is determined by the electronic structure rather than these two common mechanisms, competing exchange interactions, or a significant DMI.

Thus, the role of the DMI in ScFeGe has not been resolved in our measurements or our calculations, thus far. This is important, since materials with hexagonal symmetry and significant DMI interaction have yielded helical magnets with magnetic solitonic states and unusual magnetic domain structures that evolve with field\cite{Togawa2012,Karna2019}. Coupled to this unanswered question is the question of topological protection in a polar non-centrosymmetric helimagnet. In this case, the magnetic ordering breaks mirror symmetry, creating the possibility of topologically relevant solitonic states that persist in both fields and thermal fluctuations, and with domains of both left- and right-handed helical magnetic ordering allowed by symmetry. The data presented here suggest that ScFeGe is a weakly itinerant magnet with interesting coupling between the charge and magnetic degrees of freedom that make it an ideal material to investigate domain walls and changes to the domain structures with application of a magnetic field.

\textbf{Acknowledgements}
The experimental material presented here is supported by the U.S. Department of Energy under EPSCoR Grant No. DE-SC0012432 with additional support from the Louisiana Board of Regents. This research used resources at the High Flux Isotope Reactor, a DOE Office of Science User Facility operated by the Oak Ridge National Laboratory. Part of this work was performed at the Swiss Spallation Neutron Source SINQ, Paul Scherrer Institut, Villigen, Switzerland. The computational work conducted by W. A. S. and D. T. was also supported by the U.S. Department of Energy under EPSCoR Grant No. DE-SC0012432 with additional support from the Louisiana Board of Regents. I. V. acknowledges support from NSF Grant DMR 1410741 for theoretical work. Part of this work was performed using supercomputing resources provided by the Center for Computation and Technology (CCT) at Louisiana State University and the Center for Computational Innovations (CCI) at Rensselaer Polytechnic Institute. 

$\hspace{0.1cm}$

$\dagger$ Los Alamos National Laboratory, Los Alamos, Mail Stop E574, Los Alamos, NM 87545, USA

$\ddagger$ Department of Physics, Indiana University-Purdue University Indianapolis, Indianapolis, IN 46202, USA

$\ast$ skkarna@nsu.edu; jfditusa@iu.edu

\end{document}